\documentclass[journal,onecolumn]{IEEEtran}%
\usepackage{eurosym}
\usepackage{amsmath}
\usepackage{amsthm}
\usepackage{amsfonts}
\usepackage{amssymb}
\usepackage{graphicx}
\usepackage{tikz}
\usepackage{bm}
\usepackage{epstopdf}
\usepackage{booktabs}%
\providecommand{\U}[1]{\protect\rule{.1in}{.1in}}
\newcommand{\CLASSINPUTinnersidemargin}{3cm}
\newcommand{\CLASSINPUToutersidemargin}{3cm}

\theoremstyle{plain}
\newtheorem{theorem}{Theorem}

\theoremstyle{definition}

\newtheorem{example}{Example}
\newtheorem{remark}{Remark}

\newlength{\drop}

\newcommand*{\titleUL}
{\begingroup
\drop=0.1\textheight
\vspace*{0.5\drop}
\begin{center}
\rule{\textwidth}{1pt}\par
\vspace{0.4\baselineskip}
{\huge\bfseries  Spline Waveforms and Interference Analysis for 5G \\ \vspace*{0.4cm} Random Access with Short Message Support 
}\\[0.5\baselineskip]
\rule{\textwidth}{1pt}\par
\vfill
{\Large\textsc{Gerhard Wunder\IEEEauthorrefmark{1}, Martin Kasparick\IEEEauthorrefmark{2}, Peter Jung\IEEEauthorrefmark{3}}}
\vfill
\IEEEauthorrefmark{1}Freie Universit\"at Berlin\footnote{Gerhard Wunder is also with the Fraunhofer Heinrich-Hertz-Institut Berlin.} (\textit{g.wunder@fu-berlin.de})\\
\IEEEauthorrefmark{2}Fraunhofer Heinrich-Hertz-Institut, Berlin (\textit{martin.kasparick@hhi.fraunhofer.de})\\
\IEEEauthorrefmark{3}Technische Unversit{\"a}t Berlin (\textit{peter.jung@tu-berlin.de})
\vspace{1cm}
\begin{abstract}
One of the main drivers for new waveforms in future 5G
wireless communication systems is to handle efficiently the variety of traffic types and requirements.
In this paper, we introduce a new random access within the standard acquisition procedures to
support sporadic traffic as an enabler of the Internet of Things (IoT). The major challenge hereby is to
cope with the highly asynchronous access of different devices and to allow transmission
of control signaling and payload "in one shot". We address this challenge by using a waveform design approach
based on bi-orthogonal frequency division multiplexing where transmit orthogonality is replaced in favor of better temporal and spectral properties. We show that this approach allows data transmission in frequencies that otherwise have to remain unused. More precisely, we utilize frequencies previously used as guard bands, located towards the standard synchronous communication pipes as well as in between the typically small amount of resources used by each IoT device. We demonstrate the superiority of this waveform approach over the conventional random access using a novel mathematical approach and numerical experiments.
\footnotetext[2]{This work was carried out within the 5GNOW project, supported by the
European Commission within FP7 under grant 318555, and within DFG grant
JU-2795/2.}
\footnotetext[3]{Parts of this work were presented at the European Wireless Conference 2014 \cite{Kasparick_EW14}.}
\end{abstract}
\vfill
\vfill
\textbf{Keywords}: Waveform design, bi-orthogonal frequency division multiplexing, splines, MTC, IoT, massive random access
\vfill
\end{center}
{\itshape This article appeared in parts in 
\begin{enumerate}
 \item  New Physical Layer Waveforms for 5G, book chapter in “Towards 5G: Applications, Requirements \& Candidate Technologies”, Editors: R.Vannithamby and S. Talwar, John Wiley \& Sons Ltd, 2016 
 \item  New Waveforms for New Services in 5G, book chapter in “Orthogonal Waveforms and Filter Banks for Future Communication Systems”,  Editors: M. Renfors, X. Mestre, E. Kofidis, and F. Baader, Elsevier Academic Press, to appear 2017.
\end{enumerate}
}
\endgroup}

\graphicspath{{./figures/}}
\begin{document}

\addtolength{\baselineskip}{0.2cm}

\pagestyle{empty}\titleUL\thispagestyle{empty}

\section{Introduction}

The Internet of Things (IoT) is expected to foster the development of 5G
wireless networks and requires efficient access of sporadic traffic generating
devices. Such devices are most of the time inactive but regularly access the
Internet for minor/incremental updates with no human interaction, e.g.,
machine-type-communication (MTC). Sporadic traffic will dramatically increase
in the 5G market and, obviously, such traffic should not be forced to be
integrated into the bulky synchronization procedure of current 4G cellular
systems \cite{ComMag5GNOW,LTE_SIG}.

The new conceptional approach in this paper is to use an extended physical
layer random access channel (PRACH), which achieves device acquisition and
(possibly small) payload transmission "in one shot". Similar to the
implementation in UMTS, the goal is to transmit small user data packets using
the PRACH, without maintaining a continuous connection. So far, this is not
possible in LTE, where data is only carried using the physical uplink shared
channel (PUSCH) so that the resulting control signaling effort renders
scalable sporadic traffic (e.g., several hundred nodes in the cell)
infeasible. By contrast, in our design a data section is introduced between
synchronous PUSCH and standard PRACH, called \textit{D-PRACH} (\emph{Data
PRACH}) supporting asynchronous data transmission \cite{Kasparick_EW14}.
Clearly, by doing so, sporadic traffic is removed from standard uplink data
pipes resulting in drastically reduced signaling overhead and complexity. In
addition, this would improve operational capabilities and network performance
as well as user experience and life time of autonomous MTC nodes
\cite{ComMag5GNOW,LTE_SIG}. Waveform design in this context is a very timely
and important topic \cite{Schaich_VTC14-Spring,Berg_DSD14,Kasparick_EW14}. Of
particular importance is also the line of work in the EU projects METIS
(www.metis2020.eu) and 5GNOW (www.5gnow.eu).

We assume that each D-PRACH's data resource contains only a very few number of
subcarriers (about 5-20 subcarriers). In addition, in a 5G system, we can expect
that there is a massive number of MTC devices, which will concurrently employ
these data resources in an uncoordinated fashion. In the simplest approach,
the D-PRACH uses the guard bands between PRACH and PUSCH, which is the focus of
this paper\footnote{Notably, in an extended setting this region can be
enlarged (by higher layer parameters) but, clearly, at some point new
efficient channel estimation must be devised different to the proposal in this
paper. Recent results in METIS and 5GNOW have outlined a sparse signal
processing approach to cope with this situation \cite{RACH_ICC14,RACH_Metis14}%
.}. We show that waveform design in such a setting is necessary since the OFDM
waveform used in LTE cannot handle the highly asynchronous access of different
devices with possible negative delays or delays beyond the cyclic prefix (CP).
Clearly, guards could be introduced between the individual (small) data
sections and to the PUSCH which, though, makes the approach again very
inefficient. \emph{Our results indeed show that up to four subcarriers can be
obtained compared to a standard 4G OFDM setting}.

For waveform design, we propose a bi-orthogonal frequency division multiplexing
(BFDM) based approach where we replace orthogonality of the set of transmit
and receive pulses with bi-orthogonality, i.e., they are pairwise (not
individually) orthogonal. Thus, there is more flexibility in designing the
transmit prototype pulse (or waveform), e.g., in terms of side-lobe
suppression and robustness to time and frequency asynchronisms
\cite{kozek_98_jsac}. The BFDM approach together with a suitable waveform is
well suited to sporadic traffic, since the PRACH symbols are relatively long
so that transmission is very robust to (even negative) time offsets. In
addition, BFDM is also more robust to frequency offsets in the transmission,
which, as well-known, typically sets a limit to the symbol duration in OFDM
transmission. Finally, the concatenation of BFDM and OFDM symbols together in
a frame requires a good tail behavior of the transmit pulse in order to keep
the distortion to the payload carrying subcarriers in PUSCH small. The
excellent and controllable tradeoff between performance degradation due to
time and frequency offsets is the main advantage of BFDM with respect to
conventional OFDM.

We investigate the performance of the proposed approach using mathematical
analysis and numerical experiments where, for comparison, a standard LTE
system serves as a baseline. A part of the numerical results were already presented 
in \cite{Kasparick_EW14}. We show how the new approach can actually reduce
the interference to the PUSCH region, experienced in particular by users whose
resources are close to the PRACH. Moreover, we demonstrate that the
performance in the new D-PRACH region is significantly improved by the
pulse-shaping approach when multiple, completely asynchronous, users transmit
data in adjacent frequency bands.

\subsection{System model}

\label{sec:syssetup}

We consider a simple uplink model of a single cell network, where each mobile
station and the base station are equipped with a single antenna. We assume
there exist two channels --in LTE terminology-- the PUSCH and the PRACH. In
the PUSCH, the data bearing signals are transmitted from synchronized users to
the base station using SC-FDMA. A small part of the resources is reserved for
PRACH, in which, at the first step of the RACH procedure, users send preambles
that contain unique signatures. In this paper, we mainly deal with the PRACH
design, trying to leave PUSCH operations as unaffected as possible. Specific
system parameters can be found in Table \ref{Tab:spec} and standard textbooks
\cite{Sesia:2009}.

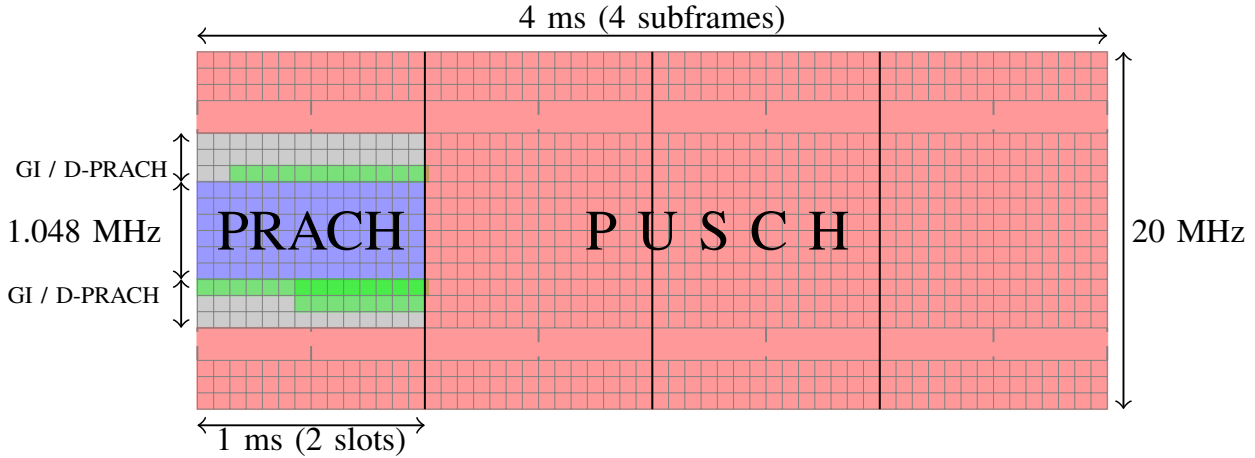
\begin{figure}[ptb]
\centering\begin{minipage}{\linewidth}
\centering
\scalebox{1.8}{\begin{tikzpicture}[scale=.12]\footnotesize
 
\begin{scope}[thick]   

  \foreach \x in {16}{
  \filldraw[thin,blue,opacity=.4] (2-\x, 13-.5*\x)
    rectangle (8-.5*\x,11);

  \filldraw[thin,gray,opacity=.4] (2-\x, 19-.5*\x)
    rectangle (8-.5*\x,14);

  \filldraw[thin,gray,opacity=.4] (2-\x, 10-.5*\x)
    rectangle (8-.5*\x,5);

   \filldraw[thin,green,opacity=.4] (2-\x, 12-.5*\x)
     rectangle (13-.8*\x,5);

  \filldraw[thin,green,opacity=.4] (8-\x, 11-.5*\x)
    rectangle (8-.5*\x,5);

  \filldraw[thin,green,opacity=.4] (4-\x, 19-.5*\x)
    rectangle (13-.8*\x,12);

    \filldraw[thin,red,opacity=.4] (2-\x, 5-.5*\x)
    rectangle (8-.5*\x, 2);    
  
  \filldraw[thin,red,opacity=.4] (2-\x, 22-.5*\x)
    rectangle (8-.5*\x, 19); 
    
     \filldraw[thin,red,opacity=.4] (30-\x, 5-.5*\x)
    rectangle (8-.5*\x, 19); 
    
    \filldraw[thin,red,opacity=.4] (44-\x, 5-.5*\x)
    rectangle (22-.5*\x, 19); 
    
    \filldraw[thin,red,opacity=.4] (58-\x, 5-.5*\x)
    rectangle (36-.5*\x, 19);

 }  
\end{scope}

\draw [help lines, step=1cm] (-14,2) grid (42,14);
\draw [help lines, step=1cm] (-14,16) grid (42,19);
\draw [help lines, step=1cm] (-14,-3) grid (42,0);

\draw [-, dashed, gray](-7,16) -- (-7,14);
\draw [-, dashed, gray](-14,16) -- (-14,14);
\draw [-, dashed, gray](42,16) -- (42,14);
\draw [-, dashed, gray](7,2) -- (7,0);
\draw [-, dashed, gray](21,16) -- (21,14);
\draw [-, dashed, gray](7,16) -- (7,14);

\draw [-, dashed, gray](35,16) -- (35,14);
\draw [-, dashed, gray](35,2) -- (35,0);
\draw [-, dashed, gray](21,2) -- (21,0);

\draw [-, dashed, gray](-7,0) -- (-7,2);
\draw [-, dashed, gray](-14,0) -- (-14,2);
\draw [-, dashed, gray](42,0) -- (42,2);

\draw[<->] (-14,-4) -- (0,-4) ; 
\draw [<->] (-15,11) -- (-15,5);
\draw [<->] (-15,14) -- (-15,11);
\draw [<->] (-15,5) -- (-15,2);
\draw [<->] (43,19) -- (43,-3);
\draw[<->] (-14,20) -- (42,20) ; 

\node at (14,21) {\scriptsize 4 ms (4 subframes)};
\node at (-7,-5) {\scriptsize 1 ms (2 slots)};
\node at (-21,8) {\scriptsize  1.048 MHz};
\node at (-20.5,11.8) {\tiny  GI / D-PRACH};
\node at (47,8) {\scriptsize 20 MHz};
\node at (-21,4) {\tiny  GI / D-PRACH};
\node at (-7,8) {\large PRACH};

\node at (18,8) {\large P       U       S       C       H};
\draw (0,19) -- (0,-3);
\draw (14,19) -- (14,-3);
\draw (28,19) -- (28,-3);
\end{tikzpicture}}
\end{minipage}\caption{PRACH (blue) and PUSCH (red) regions. A guard interval
(GI) separates PUSCH from PRACH in LTE (gray). Parts of this area are used for
data transmissions of asynchronous users (indicated in green) in a novel D-PRACH, whose
size can be variably determined by MAC. Rows in this illustration do not
represent true subcarrier quantities.}%
\label{fig:region}%
\end{figure}

The time-frequency resource grid for the described channels is illustrated in
Figure \ref{fig:region}. To minimize the interference between the channels,
several subcarriers on both sides of the PRACH are usually unused and serve as
a guard band to PUSCH. In this paper, however, the D-PRACH is located here,
i.e., some users represented by specific signatures use this region to send
data by sharing the small number of available subcarriers. Naturally, these
users may be completely asynchronous to either PUSCH or to each other, which is
a serious challenge for OFDM and will be handled by the BFDM approach using
the so-called "spline" waveform, which has "good" localization properties in
time and frequency.

\begin{figure}[ptb]
\centering\includegraphics[width=0.9\linewidth]{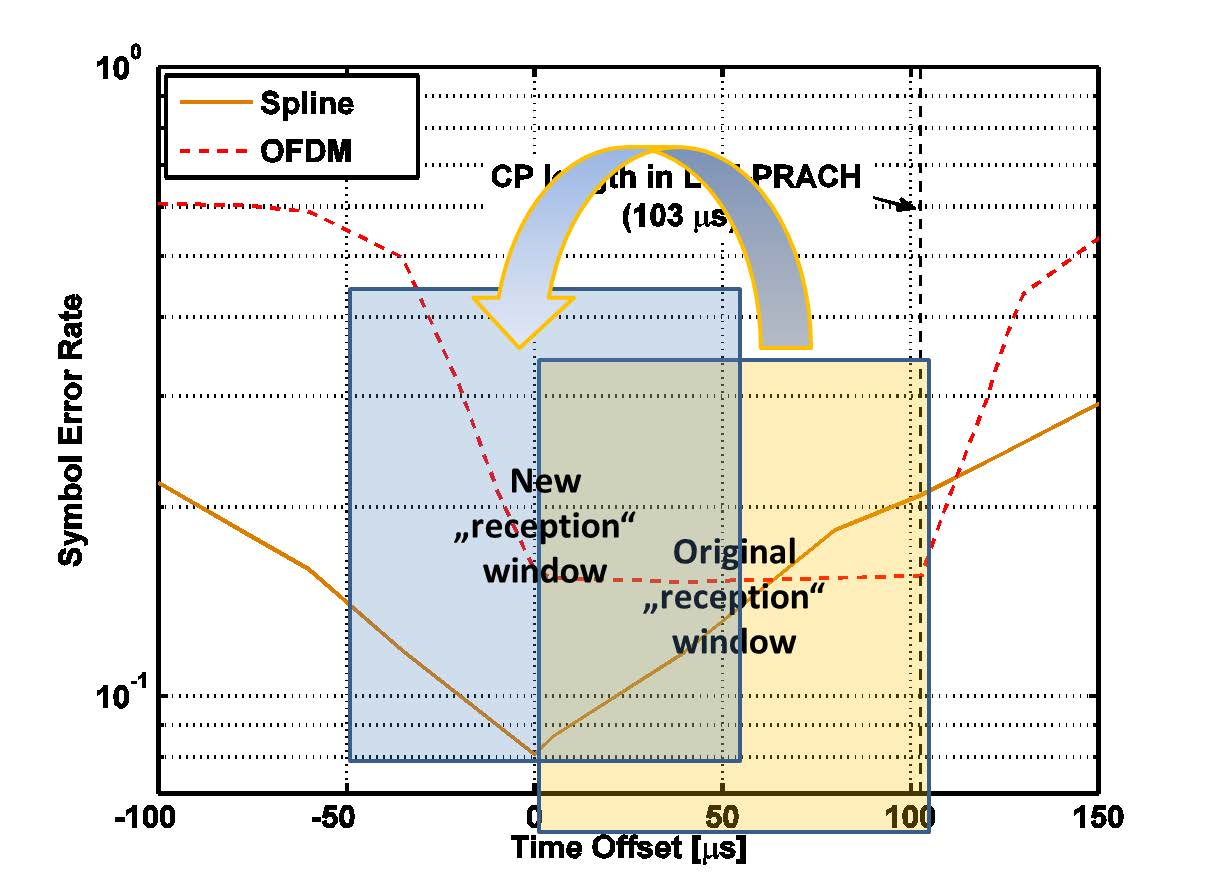}
\caption{Illustration of autonomous timing advance (ATA). The spline waveform
is in fact the candidate solution discussed in this paper and appears 
more robust to \textbf{both symmetric time and frequency offsets.} The curves depicted here 
are taken from Figure \ref{fig:simulTimeFreqOffsPRACH} in Section \ref{sec:Simulation_time_and_freq_offs} and constitute
one of our main results.}%
\label{fig:ATA}%
\end{figure}

\begin{remark}
An important property of the system is that each user can use some autonomous
timing advance (ATA), introduced in
\cite{ATA_ISWCS14},  with respect to the LTE broadcast signal. This will lead to possibly negative as well as positive
time delays of each user with respect to the receiver's "reception" window. We
will see that this can significantly lower the distortion with the new
waveforms since the distortion is "shifted" symmetric around the zero, where it
is much lower. The principle is shown in Figure \ref{fig:ATA}.

Eventually, it is worth emphasizing that the guard bands in 4G LTE are
relatively large so that the application of ATA is restricted to relatively
demanding settings with large time and frequency shifts. However, future 5G
systems are expected to have shorter symbol lengths so that the results of
this paper are applicable to much less demanding scenarios.
\end{remark}

\subsection{Organization}

The paper is organized as follows. In Section \ref{sec:Gabor}, we describe the
fundamentals of BFDM using Gabor theory and introduce the spline pulses. 
Then, in Section \ref{sec:interference}, we present our approach
to multiuser interference analysis in the context of highly asynchronous
access and provide examples for the OFDM and spline waveforms. In Section
\ref{sec:implementation}, we deal with practical implementation issues for
BFDM. In Section \ref{sec:simulations}, we investigate the performance of the
proposed approach numerically and compare to standard LTE. In Section
\ref{sec:conclusions}, we summarize our findings and draw some important conclusions.

\section{BFDM System Design}

\label{sec:Gabor}

\subsection{Elements of Gabor signaling}

Conventional OFDM and BFDM can be formulated as a pulse-shaped Gabor
multicarrier scheme. For the time-frequency multiplexing we will adopt a
two-dimensional index notation $n=(n_{1},n_{2})\in\mathbb{Z}^{2}$. 
Let $i$ denote the imaginary unit and $\mu=(\mu_{1},\mu_{2})$. 
The baseband transmit signal is then:
\begin{equation}
s(t)=\sum_{n\in{\mathcal{I}}}x_{n}\gamma_{n}(t)=\sum_{n\in{\mathcal{I}}}%
x_{n}({\boldsymbol{S}}_{\Lambda n}\,\gamma)(t), \label{eq:txsignal}%
\end{equation}
where $({\boldsymbol{S}}_{\mu}\,\gamma)(t):=e^{i2\pi\mu_{2}t}\gamma(t-\mu
_{1})$ is a
time-frequency shifted version of the transmit pulse $\gamma$, i.e., 
$\gamma_{n}:={\boldsymbol{S}}_{\Lambda n}\,\gamma$ is shifted according to a
lattice $\Lambda\mathbb{Z}^{2}$. The lattice is generated by the $2\times2$
real generator matrix $\Lambda$ and the indices $n=(n_{1},n_{2})$ range over
the doubly-countable set ${\mathcal{I}}\subset\mathbb{Z}^{2}$, referring to
the data burst to be transmitted. The coefficients $x_{n}$ are the random
complex data symbols at time instant $n_{1}$ and subcarrier index $n_{2}$ with
the property $\mathbb{E}\{\boldsymbol{x}\boldsymbol{x}^{\ast}\}={\mathbb{I}}$
(from now on $\,\bar{\cdot}$ always denotes complex conjugate, $\cdot
^{\ast}$ means conjugate transpose, and $\boldsymbol{x}=(\dots,x_{n}%
,\dots)^{T}$). We will denote the linear time-variant channel by the operator
$\boldsymbol{{\mathcal{H}}}$ and by $n(t)$ an additive distortion (a
realization of a noise process).

In practice, $\Lambda$ is usually diagonal, i.e., $\Lambda=\text{diag}(T,F)$ and
the time-frequency sampling density is related to the bandwidth efficiency (in
complex symbols) of the signaling, i.e., $\epsilon:=|\det\Lambda^{-1}%
|=(TF)^{-1}$. The received signal is:
\begin{equation}
r(t)=(\boldsymbol{{\mathcal{H}}}s)(t)+n(t)=\int_{\mathbb{R}^{2}}%
\boldsymbol{\Sigma}(\mu)({\boldsymbol{S}}_{\mu}s)(t)d\mu+n(t),
\end{equation}
with $\boldsymbol{\Sigma}:\mathbb{R}^{2}\rightarrow\mathbb{C}$ being a
realization of the (causal) channel spreading function of finite support. In
the wide-sense stationary uncorrelated scattering (WSSUS) assumption
\cite{bello:wssus}, the channel statistics is characterized by the second order
statistics of $\boldsymbol{\Sigma}$, given as the scattering function
$\boldsymbol{C}:\mathbb{R}^{2}\rightarrow\mathbb{R}_{+}$:
\begin{equation}
\mathbb{E}\{\boldsymbol{\Sigma}(\mu)\overline{\boldsymbol{\Sigma}(\mu^{\prime
})}\}=\boldsymbol{C}(\mu)\delta(\mu-\mu^{\prime}). \label{eq:wssus:assumptions}%
\end{equation}
Moreover, we assume $\mathbb{E}\{\boldsymbol{\Sigma}(\mu)\}=0$ and
$\lVert\boldsymbol{C}\rVert_{1}=1$. To obtain the (unequalized) symbol
$\tilde{x}_{m}$ on time-frequency slot $m\in{\mathcal{I}}$, the receiver
projects on $g_{m}:={\boldsymbol{S}}_{\Lambda m}g$:%
\begin{equation}
\tilde{x}_{m}:=\langle g_{m},r\rangle=\int\,e^{-i2\pi(\Lambda m)_{2}%
t}\overline{g(t-(\Lambda m)_{1})}\,r(t)dt,
\end{equation}
using the ${\mathcal{L}}_{2}$ scalar product $\langle\cdot,\cdot
\rangle:=\langle\cdot,\cdot\rangle_{{\mathcal{L}}_{2}}$. By introducing the
elements
\begin{equation}
H_{m,n}:=\langle g_{m},\boldsymbol{{\mathcal{H}}}\gamma_{n}\rangle
=\int_{\mathbb{R}^{2}}\Sigma(\mu)\langle g_{m},{\boldsymbol{S}}_{\mu}%
\gamma_{n}\rangle d\mu
\end{equation}
of the channel matrix $H\in\mathbb{C}^{|{\mathcal{I}}|\times|{\mathcal{I}}|}$,
the overall transmission can be formulated as a system of linear equations
$\tilde{\boldsymbol{x}}=\boldsymbol{H}\boldsymbol{x}+\tilde{\boldsymbol{n}}$, 
where $\tilde{\boldsymbol{n}}=(\dots,\langle g_{m},n\rangle,\dots)^{T}$ is the
vector of the projected noise. We use the AWGN assumption such that
$\boldsymbol{\tilde{n}}$ is Gaussian random vector with independent
components, each having variance ${\sigma^{2}}:=\mathbb{E}\{|r_{m}%
,n\rangle|^{2}\}$. The diagonal elements
\begin{equation}
H_{m,m}=\int_{\mathbb{R}^{2}}\Sigma(\mu)e^{i2\pi(\mu_{1}(\lambda m)_{2}%
-\mu_{2}(\lambda m)_{1})}{\mathbf{A}}_{g\gamma}(\mu)d\mu.
\end{equation}
Here, 
\begin{equation}
{\mathbf{A}}_{g\gamma}(\mu):=\langle g,{\boldsymbol{S}}_{\mu}\gamma
\rangle=\int_{\mathbb{R}}g(t)\left(  {\boldsymbol{S}}_{\mu}\gamma\right)
(t)dt\label{eq:crossambiguity}%
\end{equation}
is the well known cross-ambiguity function of $g$ and $\gamma$.

\begin{example}
A lattice can be described by a so-called generator matrix, which determines
the geometry. For cp-OFDM, we can define the matrix
\begin{equation}
\Lambda=%
\begin{bmatrix}
T+T_{cp} & 0\\
0 & \frac{1}{T}%
\end{bmatrix}
.
\end{equation}
At the transmitter, the rectangular pulse
\begin{equation}
\gamma(t)=\frac{1}{\sqrt{T+T_{cp}}}\chi_{\lbrack-T_{[}[cp],T]}(t)
\end{equation}
is used, with $\chi_{\lbrack-T_{[}[cp],T]}$ being the characteristic function
of the interval $[-T_{[}[cp],T]$. The receiver obtains the complex symbol as
\begin{equation}
\tilde{x}_{m}=\int_{-\infty}^{\infty}g(t-(\Lambda n)_{1})e^{j2\pi(\Lambda
n)_{2}t}dt,
\end{equation}
using the rectangular pulse
\begin{equation}
g(t)=\frac{1}{\sqrt{T}}\chi_{\lbrack0,T]}(t)
\end{equation}
for the removal of the CP.
\end{example}

\subsection{Completeness, Localization and Gabor Frames}

In the following, we will collect some well-known statements on (bi-infinite)
Gabor families $\mathcal{G}(\gamma,\Lambda):=\{\boldsymbol{S}_{\Lambda
n}\gamma\}_{n\in\mathbb{Z}^{2}}$. A more detailed discussion of these concepts
with respect to multicarrier transmission in doubly-dispersive channels can be
found for example in \cite{jung_07_comm} and \cite{jung:wcnc08}. Based on the
generator matrix $\Lambda$, we can categorize the following regimes: we refer
to \emph{critical sampling} in case $\mathrm{det}(\Lambda)=1$, while
$\mathrm{det}(\Lambda)<1$ and $\mathrm{det}(\Lambda)>1$ is called
\emph{oversampling} and \emph{undersampling} of the time-frequency plane,
respectively. Perfect reconstruction for \emph{any} $\mathcal{I}%
\subseteq\mathbb{Z}^{2}$, i.e., $\langle g_{m},\gamma_{n}%
\rangle=\delta_{mn}$ for all $m,n\in\mathbb{Z}^{2}$, can be achieved if and
only if $\mathrm{det}(\Lambda)\geq1$. In this case, $g$ is called the dual
(bi-orthogonal) pulse to $\gamma$ with respect to the lattice $\Lambda$. If
$g=\gamma$, the Gabor family $\mathcal{G}(g,\Lambda)$ is an orthogonal basis
for its span. However, a main consequence from the \emph{Balian Low Theorem}
is, that no well-localized prototype $\gamma$ (in both time and frequency) can
generate a Gabor Riesz basis at $\mathrm{det}(\Lambda)=1$. Thus, any
orthogonal or biorthogonal signaling at the critical density is
ill-conditioned and will be highly sensitive with respect to either time or
frequency shifts. Therefore, without further constraints on the data symbols, 
one has to operate in the undersampling regime in a practical scenario.

Let us now consider the adjoint lattice, generated by $\Lambda^{\circ
}=\mathrm{det}(\Lambda)^{-1}\cdot\Lambda$, i.e., if $\mathcal{G}(\gamma
,\Lambda)$ refers to undersampling, $\mathcal{G}(\gamma,\Lambda^{\circ})$
corresponds to oversampling. An important notion here is the concept of a
\emph{Gabor frame}, i.e. $\mathcal{G}(\gamma,\Lambda^{\circ})$ establishes a
frame (for $L_{2}(\mathbb{R})$) if there are frame bounds $0<A_{\gamma}\leq
B_{\gamma}<\infty$ such that:
\begin{equation}
A_{\gamma}\lVert f\rVert_{2}^{2}\leq\sum_{n\in\mathbb{Z}^{2}}\lvert
\langle\boldsymbol{S}_{\Lambda^{\circ}n}\gamma,f\rangle\rvert^{2}\leq
B_{\gamma}\lVert f\rVert_{2}^{2} \label{eq:framecondition}%
\end{equation}
for all $f\in L_{2}(\mathbb{R})$. With $0<A_{\gamma}\leq B_{\gamma}<\infty$ we
always mean here the best bounds. The \emph{Ron-Shen Duality} now states that
the Gabor family $\mathcal{G}(\gamma,\Lambda)$ is a Riesz basis for its span
if and only if $\mathcal{G}(\gamma,\Lambda^{\circ})$ is a frame for
$L_{2}(\mathbb{R})$. Furthermore, $\mathcal{G}(g,\Lambda^{\circ})$ is a tight
frame $(A_{\gamma}=B_{\gamma})$ if and only if $\mathcal{G}(\gamma,\Lambda)$
is an orthogonal basis for its span. If $B_{\gamma}<\infty$ exists in
\eqref{eq:framecondition} for the generator matrix $\Lambda$, the sequence of
elements in $\mathcal{G}(\gamma,\Lambda)$ is called Bessel sequence. A
straightforward argument shows that $B_{\gamma}$ is the maximal eigenvalue of
the bi-infinite Gram matrix $G_{\gamma}$ with the components $(G_{\gamma
})_{m,n}=\langle\gamma_{m},\gamma_{n}\rangle$. More details on these concepts
in multicarrier transmission can be found in \cite{jung:thesis}.

\begin{example}
[{\cite[Appendix A]{jung:ieeecom:timevariant}}] The Bessel bound of cp-OFDM
signaling is given by the operator norm of $S_{\gamma}$, which is equal to the
largest eigenvalue of the Gram matrix $G_{\gamma}$, i.e., 
\begin{equation}
(G_{\gamma})_{m,n}=\delta_{m_{1},n_{1}}e^{-i\frac{\pi}{\epsilon}(n_{2}-m_{2}%
)}\frac{\sin\frac{\pi}{\epsilon}(n_{2}-m_{2})}{\frac{\pi}{\epsilon}%
(n_{2}-m_{2})}, %
\end{equation}
where $\epsilon=\frac{T_{u}}{T_{u}+T_{cp}}$. This is a Toeplitz matrix in the
frequency slots $m_{2}$ and $n_{2}$ generated by:
\begin{align}
\phi(\omega)  &  =\sum_{n=-\infty}^{\infty}e^{j\pi(2\omega-\frac{1}{\epsilon
})n}\frac{\sin\frac{\pi}{\epsilon}n}{\frac{\pi}{\epsilon}n}=1+\frac{2\epsilon
}{\pi}\sum_{n=1}^{\infty}\frac{\cos\pi(2\omega-\frac{1}{\epsilon})n\cdot
\sin\frac{\pi}{\epsilon}n}{n}\\
&  =1+\epsilon-\epsilon\left[  \left(  \frac{1}{\epsilon}-\omega\right)
\mathrm{mod}1+\omega\right]  =\epsilon\left(  \lfloor\frac{1}{\epsilon}%
-\omega\rfloor+1\right).
\end{align}
Using this, the Bessel bound for cp-OFDM can be upperbounded by $B_{\gamma}%
=2$. However, of course this is only the worst-case estimation.
Interestingly, by contrast, the Bessel bound $B_{g}$ for the receive "frame"
is exactly unity, i.e., $B_{g}=1$.
\end{example}

\subsection{Spline-based Gabor Signalling}

\begin{figure}[ptb]
\centering\includegraphics[width=0.9\linewidth]{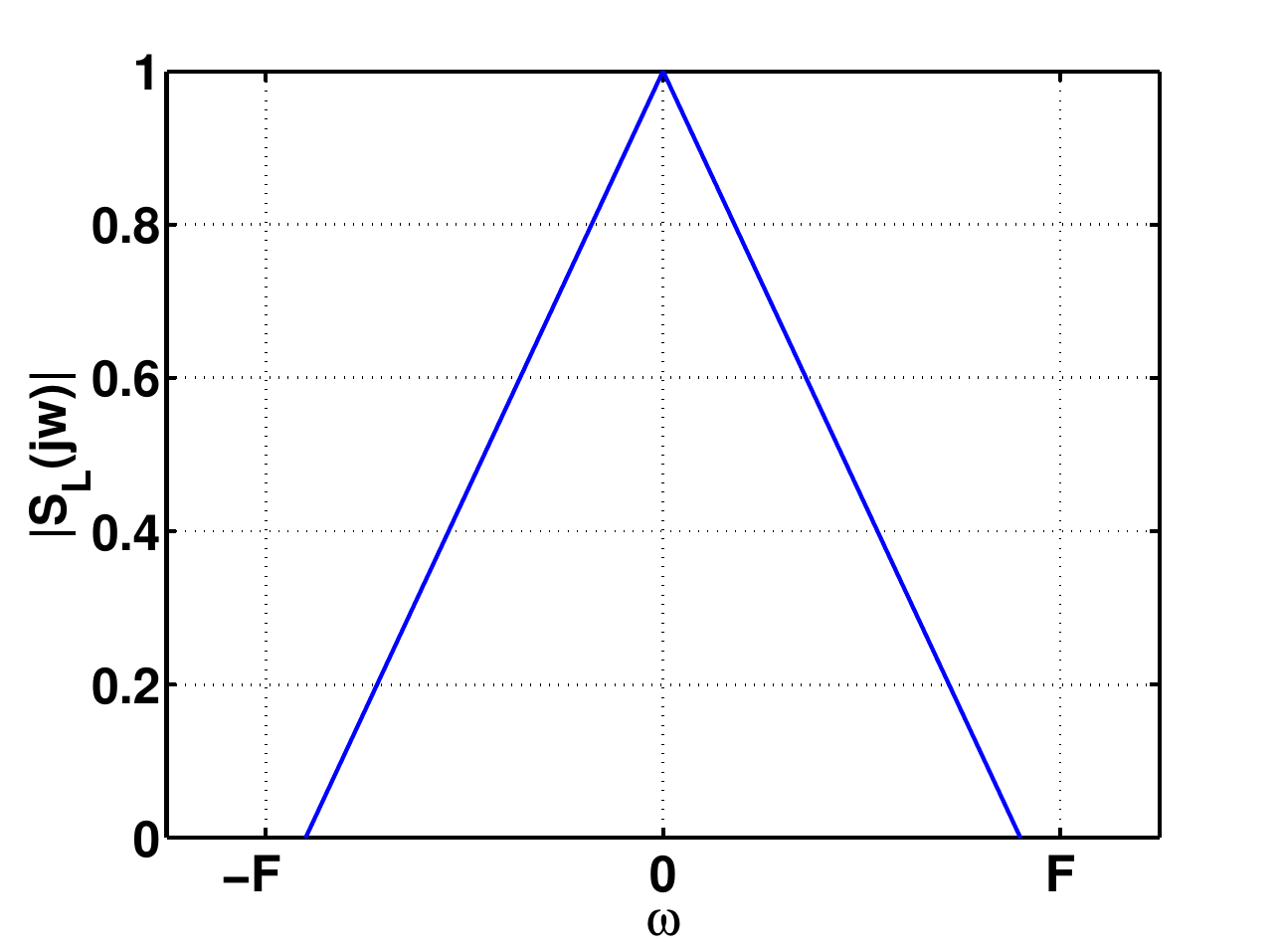}
\caption{Illustration of 1$^{st}$ order spline function. It is the result of a
convolution of two 0$^{th}$ order rectangular pulses.}%
\label{fig:spline}%
\end{figure}

As mentioned before, the used pulses $g$ and $\gamma$ play a key role and
should therefore be carefully designed. As we consider a BFDM approach, we
generate the transmit pulse $g$ according to system requirements and compute
the receive pulse $\gamma$ as the canonical dual (biorthogonal) pulse. For
this, we use a method already applied, for example, in \cite{jung_07_comm}.
Briefly explained, bi-orthogonality in a stable sense means that $g$ should
generate a Gabor Riesz basis and $\gamma$ generates the corresponding dual
Gabor Riesz basis. From the Ron-Shen duality principle \cite{ronshen:duality}
follows that $\gamma$ has the desired property if it generates a Gabor
(Weyl-Heisenberg) frame on the so called adjoint time-frequency lattice and
that frame is dual to the frame generated by $g$. This can be achieved with
the $S^{-1}$-trick explained in \cite{daubechies:tenlectures}. Side effects
such as spectral regrowth due to periodic setting when calculating the
bi-orthogonal pulses are negligible.

As a rough and well-known guideline for well-conditioning of this procedure,
the ratio of the time and frequency pulse widths (variances) $\sigma_{t}$ and
$\sigma_{f}$ should be approximately matched to the time-frequency grid ratio
\begin{equation}
\frac{T}{F}\approx\sqrt{\frac{\sigma_{t}}{\sigma_{f}}}, \label{eq:scaling}%
\end{equation}
and this should also be in the order of the channel's dispersion ratio
\cite{jung_07_comm}. However, since we focus on a design being close to the conventional LTE PUSCH and PRACH, 
of this rule, we consider only \eqref{eq:scaling} here.

We propose to construct the pulse $g$ based on the B-splines in the frequency
domain, see Figure \ref{fig:spline}. $B$-splines have been investigated in the
Gabor (Weyl-Heisenberg) setting for example in \cite{Prete:1999}. The main
reason for using the B-spline pulses is that convolution of such pulses have
excellent tail properties with respect to the $L_{1}$-norm, which is
beneficial with respect to the overlap of PRACH to the PUSCH symbols. We also
believe that they trade off well the time offset for the frequency offset
performance degradation but this is part of further on-going investigations
and beyond the conceptional approach here. Because of its fast decay in time,
we choose a second order B-spline (the \textquotedblright
tent\textquotedblright-function) in frequency domain, given by
\begin{align}
B_{2}(f)  &  =B_{1}(f)\ast B_{1}(f),\;\text{where}\\
B_{1}(f)  &  :=\chi_{\lbrack-\frac{1}{2},\frac{1}{2}]}(f)
\end{align}
(and $\ast$ denotes convolution). It has been shown in \cite{Prete:1999} that
$B_{2}(f)$ generates a Gabor frame for the $(a,b)$-grid (translating $B_{2}$
on $a\mathbb{Z}$ and its Fourier transform on $b\mathbb{Z}$) if (due to its
compact support) $a<2$ and $b\leq1/2$, and fails to be frame in the region:
\begin{equation}
\{a\geq2,\,b>0\}\cup\{a>0,1<b\in\mathbb{N}\}.
\end{equation}
Recall, that by Ron-Shen duality \cite{ronshen:duality} it follows that the
same pulse prototype $B_{2}(f)$ generates a Riesz basis on the adjoint
$(\frac{1}{b},\frac{1}{a})$-grid. In our setting, we will effectively translate
the frequency domain pulse $B_{2}(f)$ by half of its support, which corresponds
to $\frac{1}{b}=1$, and we will use $\frac{1}{b}\cdot\frac{1}{a}=\frac{5}%
{4}=1.25$ (see here also Table \ref{Tab:spec}) such that $a=\frac{4}{5}$. Therefore, it
follows that our operation point $(a,b)=(\frac{4}{5},1)$ is not in
any of two explicit $(a,b)$-regions given above. But for $1.1\leq a\leq1.9$ a
further estimate has been computed explicitly for $B_{2}(f)$ \cite[Table 2.3
on p.560]{Prete:1999}, ensuring the Gabor frame property up to $b\leq1/a$.
Finally, we like to mention that for $ab\leq1/2$ the dual prototypes can be
expressed again as finite linear combinations of $B$-splines, i.e., explicit
formulas exists in this case \cite{Laugesen2009}. Note that we can choose a
larger grid in the frequency domain and set $b\leq1/2$ so that $a=8/5$, which
is a frame (so that the spline is a Riesz basis). Hence, due to the spectral
efficiency constraint with increasing $1/b$, we also decrease the time domain
grid such that, necessarily, at some point $a\geq2$ (in the dual domain) and
so we do not get a Riesz basis (or a frame in the dual domain).

In practice, $g$ has to be of finite duration, i.e., the transmit pulse in time
domain will be smoothly truncated:
\begin{equation}
g(t)=\left(  \frac{\mathrm{sin}(B\pi t)}{B\pi t}\right)  ^{2}\chi_{\lbrack
c,d]}(t), \label{eq:bspline:truncated}%
\end{equation}
where $B$ is chosen equal to $F$ and parameters $c$ and $d$ align the pulse
within the transmission frame.
Theoretically, a (smooth) truncation in \eqref{eq:bspline:truncated} would
imply again a limitation on the maximal frequency spacing $B$
\cite{Christensen:2012}. Although the finite setting is used in our
application, the frame condition (and therefore the Riesz basis condition) is
a desired feature since it will asymptotically ensure the stability of the
computation of the dual pulse $\gamma$ and its smoothness properties.

\begin{figure}[ptb]
\centering\includegraphics[width=0.9\linewidth]{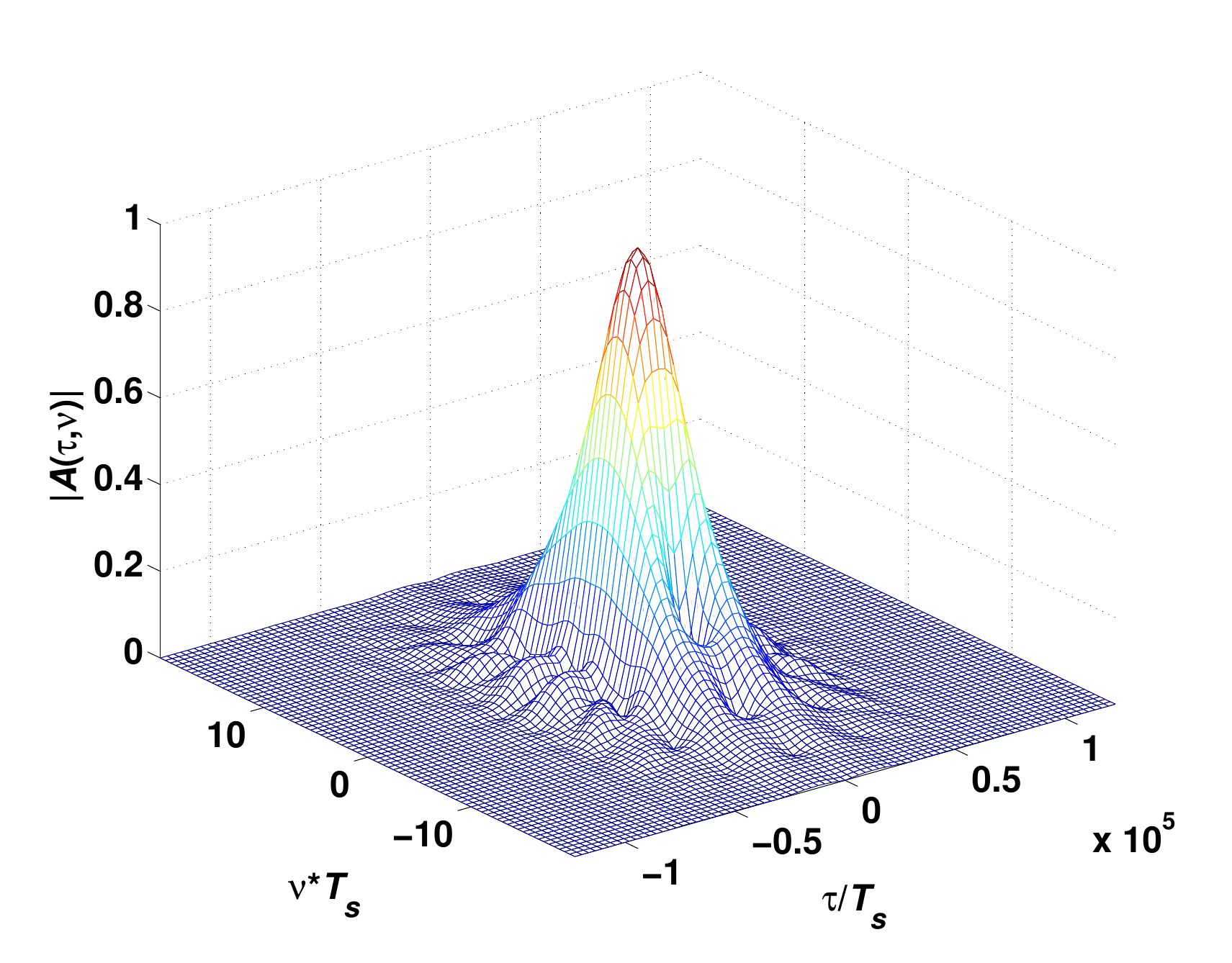}
\caption{The cross--ambiguity function $A_{g,\gamma}(\nu,\tau)$ for transmit
and receive pulses on the frame length of $4\,ms$.}%
\label{fig:ambig_fun}%
\end{figure}

To observe the pulse's properties regarding time-frequency distortions, we
depict in Figure \ref{fig:ambig_fun} the discrete cross-ambiguity function
$A_{g,\gamma}$ between pulse $g$ and $\gamma$.

 It can be observed that its value at the neighboring symbol is
already far below $10^{-3}$. Obviously, the bi-orthogonality condition states
\mbox{$A_{g,\gamma}(kT,lF)=\delta_{k,0}\delta_{t,0}$} and ensures perfect
symbol recovery in the absence of channel and noise. However, the sensibility
with respect to time-frequency distortions is related to the slope shape of
$A_{g,\gamma}$ around the grid points. Depending on the loading strategies for
these grid points it is possible to obtain numerically performance estimates
using, for example, the integration methods presented in \cite{jung_07_comm}.

\begin{figure}[ptb]
\centering\includegraphics[width=0.9\linewidth]{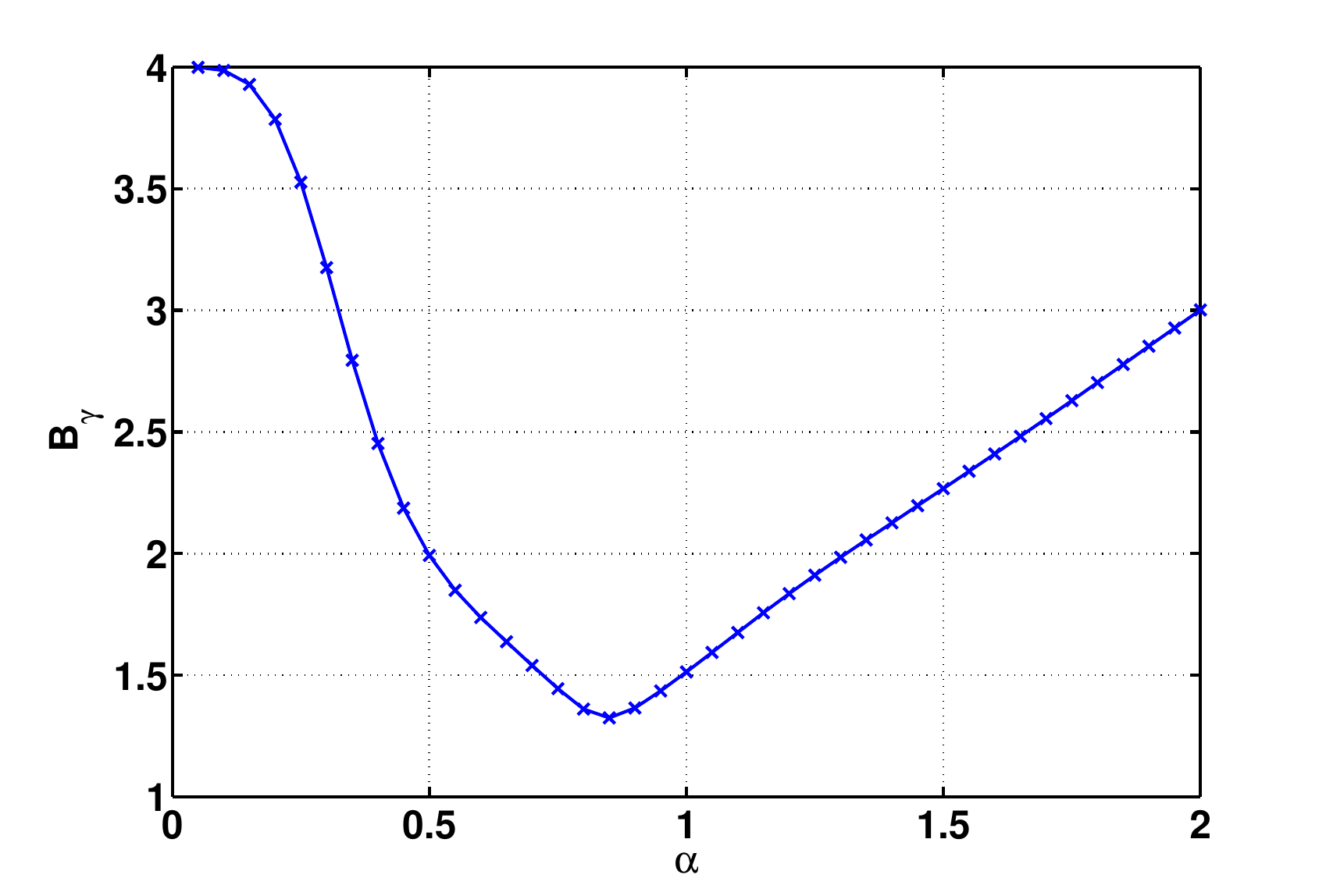}\caption{Upper
frame constant $B_{\gamma}$ over $\alpha$. A minimum of this function can be
observed around $1$ (more precisely, at $\alpha=0.85$), while for increasing
and, even more, for decreasing values of alpha the value of $B_{\gamma}$
increases.}%
\label{fig:Bgamma_Alpha}%
\end{figure}

Let us introduce a parameter $\alpha$ to scale the width of the spline pulse
in frequency. Choosing a large pulse width $\alpha$ has the
disadvantage of an increasing value of $B_{g}$ in (\ref{eq:interference_bound}%
). This is illustrated in Figure \ref{fig:Bgamma_Alpha}. It shows the upper
frame constant $B_{g}$ for the transmit pulse $g$ vs. its pulse width $\alpha
$. The frame constant is thereby calculated using the LTFAT toolbox
\cite{Sondergaard12}. It can be observed that $B_{g}$ has its smallest value
around $\alpha=0.85$, however, it never reaches the lower bound $B_{g}=1$.

\section{Multiuser Interference Analysis}

\label{sec:interference}

\subsection{A general approach}

Our system model has to capture that many users each occupy a small number of
subcarriers and each of them asynchronously (in time and frequency or both)
access this resource in an uncoordinated fashion. For a particular
time-frequency slot $m=(m_{1},m_{2})$, we will denote the (random) channel
operator as $\boldsymbol{{\mathcal{H}}}(m)$ and the asynchronism as
$\mathcal{D}\left(  m\right)  $. We assume that $\boldsymbol{{\mathcal{H}}%
}(m)$ can be estimated using channel estimation procedure while $\mathcal{D}%
\left(  m\right)  $ cannot. Writing the received complex symbol $\tilde{x}%
_{m}$ in the absence of additive noise yields
\begin{equation}
\tilde{x}_{m}=\overline{H}_{m}x_{m}+\overbrace{(H_{m,m}-\overline{H}_{m}%
)}^{\Delta_{m}}x_{m}+\underbrace{\sum_{n\in\mathcal{I},n\neq m}H_{m,n}x_{n}%
}_{\text{ICI}}%
\end{equation}
where we defined $\overline{H}_{m}:=\mathbb{E}_{|\mathcal{H}}\{H_{m,m}\}$,
i.e., the mean value conditioned on a fixed channel $\mathcal{H}$\footnote{As a
matter of fact, the expectations depends only on the marginal distribution of
$\mathcal{H}(m)$.}. Thus, the transmitted symbol $x_{m}$ will be multiplied by
a constant and disturbed by two zero mean random variables (RV), $\Delta_{m}$
and ICI. The first RV $\Delta_{m}$ represents a distortion, which comes from
the randomness of $\mathcal{D}\left(  m\right)  $, the second term ICI
represents both. The mean power of both contributions, conditioned on a fixed
channel $\boldsymbol{\mathcal{H}}$, are $D_{m}:=\mathbb{E}_{|\mathcal{H}%
}\{|\Delta_{m}|^{2}\}=P_{m}-|\overline{H}_{m}|^{2}$ where $P_{m}%
:=\mathbb{E}_{|\mathcal{H}}\{|H_{m,m}|^{2}\}$ and $I_{m}:=\mathbb{E}%
\{|\text{ICI}|^{2}\}$. Each element of the distortion sum $\sum_{n\neq
m}H_{m,n}x_{n}$ is given by:%
\begin{equation}
H_{m,n}=\langle g_{m},\mathcal{D}\left(  n\right)  \boldsymbol{{\mathcal{H}}%
}(n)\gamma_{n}\rangle.
\end{equation}

\begin{remark}
Notably, even if $\mathcal{D}\left(  n\right)  $ is the identity (synchronous
access) $\tilde{x}_{m}$ is affected by all other individual contributions
where the operators depend also on the index $n=(n_{1},n_{2})$. Hence, the
performance for individual slots will be quite different, which complicates the
situation, and no analytical approach is available so far! If all
$\mathcal{D}\left(  n\right)$ and $\boldsymbol{{\mathcal{H}}}(n)$ are independent
of $n$, standard analysis can be used \cite{jung_07_comm}.
\end{remark}

To find a tractable way, we consider the following approach: We assume that the
$H_{m,m}$ can be estimated, and we consider the distortion sum $\sum_{n\neq
m}H_{m,n}x_{n}$ averaged over all the subcarriers $m$. Obviously, this will
average out individual interference for a specific subcarrier, but we can
assume that these interference terms do not differ much. Individual performance
is then measured by $H_{m,m}$ only! Then, we average over the random operators
$\mathcal{D}\left(  n\right)  $ and $\boldsymbol{{\mathcal{H}}}(n)$. Let us
first consider the sums:%

\begin{align}
\sum_{m}\sum_{n\neq m}\left\vert H_{m,n}\right\vert ^{2}  &  =\sum_{m}%
\sum_{n\neq m}\left\vert \langle g_{m},\mathcal{D}\left(  n\right)
\boldsymbol{{\mathcal{H}}}(n)\gamma_{n}\rangle\right\vert ^{2}\newline\\
&  =\sum_{n}\sum_{m}\left\vert \langle g_{m},\mathcal{D}\left(  n\right)
\boldsymbol{{\mathcal{H}}}(n)\gamma_{n}\rangle\right\vert ^{2}-\sum
_{m}\left\vert \langle g_{m},\mathcal{D}\left(  m\right)
\boldsymbol{{\mathcal{H}}}(m)\gamma_{m}\rangle\right\vert ^{2}\newline\\
&  \leq B_{g}E_{g}\sum_{n}\left\Vert \mathcal{D}\left(  n\right)
\boldsymbol{{\mathcal{H}}}(n)\gamma_{n}\right\Vert _{2}^{2}-\sum_{m}\left\vert
\langle g_{m},\mathcal{D}\left(  m\right)  \boldsymbol{{\mathcal{H}}}%
(m)\gamma_{m}\rangle\right\vert ^{2}\newline\\
&  =B_{g}E_{g}\sum_{n}\left\Vert \mathcal{D}\left(  n\right)
\boldsymbol{{\mathcal{H}}}(n)\gamma_{n}\right\Vert _{2}^{2}-\sum_{m}\left\vert
H_{m,m}\right\vert ^{2}.%
\end{align}
Here, $B_{g}$ is the Bessel bound of the Gabor family $\mathcal{G}%
(g,\Lambda)$. In the last step, we see that only the "action" of the operators
$\{\mathcal{D}\left(  n\right)  ,\boldsymbol{{\mathcal{H}}}(n)\}$ on $\gamma$
is relevant. We have set without loss of generality $||\gamma||_{2}^{2}=1$ and
$1\leq||g||_{2}^{2}\leq E_{g}$ (typically $E_{g}\approx1$). Next, we compute
the expectations and we use $D_{m}=P_{m}-|\overline{H}_{m}|^{2}$:%

\begin{align}
&  \mathbb{E}\sum_{m}\sum_{n\neq m}\left\vert H_{m,n}\right\vert
^{2}+\mathbb{E}\sum_{m}D_{m}\newline\\
&  \leq B_{g}E_{g}\sum_{n}\mathbb{E}\left\Vert \mathcal{D}\left(  n\right)
\boldsymbol{{\mathcal{H}}}(n)\gamma_{n}\right\Vert _{2}^{2}-\sum_{m}%
\mathbb{E}\left\vert H_{m,m}\right\vert ^{2}+\mathbb{E}\sum_{m}D_{m}\newline\\
&  \leq B_{g}E_{g}\sum_{n}\mathbb{E}\left\Vert \mathcal{D}\left(  n\right)
\boldsymbol{{\mathcal{H}}}(n)\gamma_{n}\right\Vert _{2}^{2}-\mathbb{E}\sum
_{m}(P_{m}-D_{m})\newline\\
&  =B_{g}E_{g}\sum_{n}\mathbb{E}\left\Vert \mathcal{D}\left(  n\right)
\boldsymbol{{\mathcal{H}}}(n)\gamma_{n}\right\Vert _{2}^{2}-\sum_{m}%
\mathbb{E}|\overline{H}_{m}|^{2}.%
\end{align}
We assume that the asynchronisms cannot increase the received power. For the
first term, we estimate
\begin{equation}
\sum_{n}\mathbb{E}\left\Vert \mathcal{D}\left(  n\right)
\boldsymbol{{\mathcal{H}}}(n)\gamma_{n}\right\Vert _{2}^{2}\leq\sum
_{n}\mathbb{E}\lVert\boldsymbol{\mathcal{H}}\left(  n\right)  \gamma_{n}%
\rVert_{2}^{2}\leq\sum_{n}\lVert\boldsymbol{C}_{n}\rVert_{1},
\end{equation}
according to (\ref{eq:wssus:assumptions}). It remains to bound the second
term
\begin{equation}
\mathbb{E}|\overline{H}_{m}|^{2}=\mathbb{E}\left\vert |\mathbb{E}%
_{|\mathcal{H}}\langle g_{m},\mathcal{D}\left(  m\right)
\boldsymbol{{\mathcal{H}}}(m)\gamma_{m}\rangle\right\vert ^{2}%
\end{equation}
for some $m$. For $a,b\in\mathbb{R}^{2}$, we abbreviate $[a,b]:=a_{1}%
b_{2}-a_{2}b_{1}$ (the symplectic form). Define now the following function:
\begin{align}
s_{m}(\mu)  &  :=\mathbb{E}_{|\mathcal{H}}\langle g_{m},\mathcal{D}\left(
m\right)  {\boldsymbol{S}}_{\mu}\gamma_{m}\rangle\label{eq:def_s_m}\\
&  =e^{-j2\pi\lbrack\mu,\Lambda m]}\mathbb{E}_{|\mathcal{H}}\langle
g,{\boldsymbol{S}}_{\Lambda m}^{\ast}\mathcal{D}\left(  m\right)
{\boldsymbol{S}}_{\Lambda m}{\boldsymbol{S}}_{\mu}\gamma\rangle,
\end{align}
which essentially contains the distortion of the $\mu$th contribution in terms
of the pulses conjugated by $\boldsymbol{S}_{\Lambda m}$, i.e., 
\textquotedblright shifted\textquotedblright\ to TF-slot $m$ in the
time-frequency plane. For a fixed channel $\boldsymbol{\Sigma}$, we have
$\overline{H}_{m}=\langle\boldsymbol{\Sigma},s_{m}\rangle$ and on average, with
respect to $\boldsymbol{{\mathcal{H}}}(m)$, we have:
\begin{equation}
\mathbb{E}\{|\overline{H}_{m}|^{2}\}=\langle\boldsymbol{C}_{m},|s_{m}%
|^{2}\rangle.
\end{equation}
Hence, altogether we have the following upper bound on the total expected
distortion:
\begin{equation}
\mathbb{E}\sum_{m\in\mathcal{I}}(I_{m}+D_{m})\leq\sum_{m\in\mathcal{I}}%
(E_{g}B_{g}\lVert\boldsymbol{C}_{m}\rVert_{1}-\langle\boldsymbol{C}_{m}%
,|s_{m}|^{2}\rangle).
\end{equation}
Let us fix the normalization such that $\sum_{m\in\mathcal{I}}\lVert
\boldsymbol{C}_{m}\rVert_{1}/|\mathcal{I}|=1$. Hence, averaging over
$\mathcal{I}$, we have proved the following theorem.

\begin{theorem}
Suppose $\lVert\gamma_{m}\rVert_{2}^{2}=1$ (without loss of
generality)$,\lVert g_{m}\rVert_{2}^{2}=E_{g}$ such that $\langle
\boldsymbol{g}_{m},\gamma_{m}\rangle=1$ (perfect reconstruction in noiseless
case). The average distortion power per subcarrier is upperbounded by%
\begin{equation}
\frac{1}{|\mathcal{I}|}\mathbb{E}\sum_{m\in\mathcal{I}}(I_{m}+D_{m})\leq
E_{g}B_{g}-\frac{1}{|\mathcal{I}|}\sum_{m\in\mathcal{I}}\langle\boldsymbol{C}%
_{m},|s_{m}|^{2}\rangle,
\end{equation}
where:%
\[
s_{m}(\cdot)=\mathbb{E}_{|\mathcal{H}}\langle g_{m},\mathcal{D}\left(
m\right)  {\boldsymbol{S}}_{(\cdot)}\gamma_{m}\rangle.
\]

\end{theorem}

\begin{example}
As a special case, assume now a deterministic time-frequency shift
$\mathcal{D}\left(  m\right)  =\boldsymbol{S}_{\nu(m)}$. This distortion is
non-random and energy preserving, i.e., $\lVert\mathcal{D}\left(  m\right)
g_{m}\rVert_{2}=\lVert g_{m}\rVert_{2}$. Evaluating the function $s_{m}$ in
(\ref{eq:def_s_m}) gives:
\begin{align}
s_{m}(\mu(m))  &  =e^{-i2\pi\lbrack\mu(m),\Lambda m]}\langle g,{\boldsymbol{S}%
}_{\Lambda m}^{\ast}{\boldsymbol{S}}_{\nu(m)}{\boldsymbol{S}}_{\Lambda
m}{\boldsymbol{S}}_{\mu(m)}\gamma\rangle=e^{-i2\pi\lbrack\nu(m)+\mu(m),\Lambda
m]}\langle g,{\boldsymbol{S}}_{\nu(m)}{\boldsymbol{S}}_{\mu(m)}\gamma\rangle\\
&  =e^{-i2\pi\left(  \lbrack\nu(m)+\mu(m),\Lambda m]+\nu_{1}(m)\mu
_{2}(m)\right)  }\langle g,{\boldsymbol{S}}_{\nu(m)+\mu(m)}\gamma\rangle\\
&  =e^{-i2\pi\left(  \lbrack\nu(m)+\mu(m),\Lambda m]+\nu_{1}(m)\mu
_{2}(m)\right)  }{\mathbf{A}}_{g\gamma}(\nu(m)+\mu(m)).
\end{align}
Hence, in the AWGN case we have:
\begin{equation}
\frac{1}{\left\vert {\mathcal{I}}\right\vert }\mathbb{E}\sum_{m\in\mathcal{I}%
}(I_{m}+D_{m})\leq E_{g}B_{g}-\frac{1}{\left\vert {\mathcal{I}}\right\vert
}\sum_{m\in\mathcal{I}}|{\mathbf{A}}_{g\gamma}(\nu(m))|^{2}.
\label{eq:interference_bound}%
\end{equation}

\end{example}

\subsection{OFDM}

\label{sec:timeandfreqoffsets}

\begin{figure}[ptb]
\centering\includegraphics[width=0.9\linewidth]{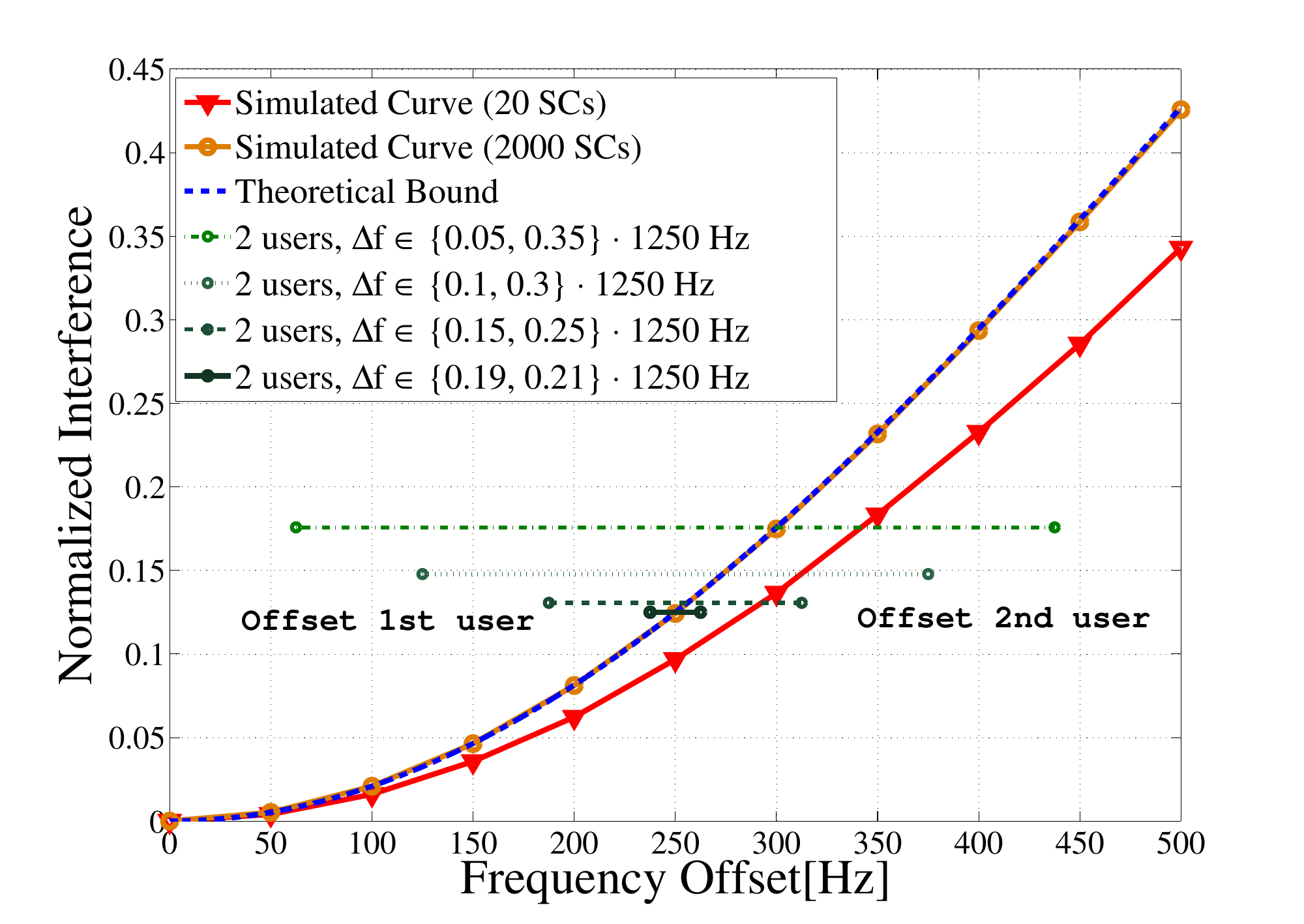}
\caption{Simulated interference over frequency offset using different numbers
of subcarriers, compared to theoretical bound. In addition, the behavior of
the two-user case is illustrated, where each user gets half of the available
subcarriers and has a different frequency offset. The (aggregate) interference
converges to the value of the bound when the differences between the two
offsets, which are centered around $0.2\cdot1250Hz$, and their average become
smaller.}%
\label{fig:IntVsFrOff_OFDM}%
\end{figure}

The cross ambiguity function for $\gamma$ and $g$, as introduced in
(\ref{eq:crossambiguity}), can be compactly written as (see
\cite{jung:ieeecom:timevariant})
\begin{align}
{\mathbf{A}}_{g\gamma}(\nu)  &  =\frac{\sin\pi\nu_{2}(T_{u}-|[\nu
_{1}]_{\text{cp}}|)}{\pi\nu_{2}T_{u}}e^{j(\phi_{0}-\pi\nu|[\nu_{1}%
]_{\text{cp}}|)}\chi_{\lbrack-T_{u},T_{u}]}([\nu_{1}]_{\text{cp}})\\
&  ={\mathbf{A}}_{gg}(([\nu_{1}]_{\text{cp}},\nu_{2})),
\label{eq:offsets:ambiguity}%
\end{align}
where the phase $\phi_{0}=\pi\nu T_{u}$ is related to our choice of time
origin and
\begin{equation}
\lbrack\cdot]_{\text{cp}}\ :\tau\rightarrow\lbrack\tau]_{\text{cp}}=%
\begin{cases}
\tau & \tau\leq0,\\
0 & 0<\tau<T_{cp},\\
\tau-T_{cp} & \tau\geq T_{cp}.%
\end{cases}
\end{equation}
The signal quality in the presence of time- and frequency shifts can now be
directly obtained from (\ref{eq:offsets:ambiguity}). Apart from $[\cdot
]_{\text{cp}}$ and $\sqrt{\epsilon}$ (the loss in mean signal amplitude due to
the CP) (\ref{eq:offsets:ambiguity}) agrees with the well known auto ambiguity
function $\mathbf{A}_{gg}$ for a rectangular pulse $g$ of width $T_{u}$.

\begin{figure}[ptb]
\centering\includegraphics[width=0.9\linewidth]{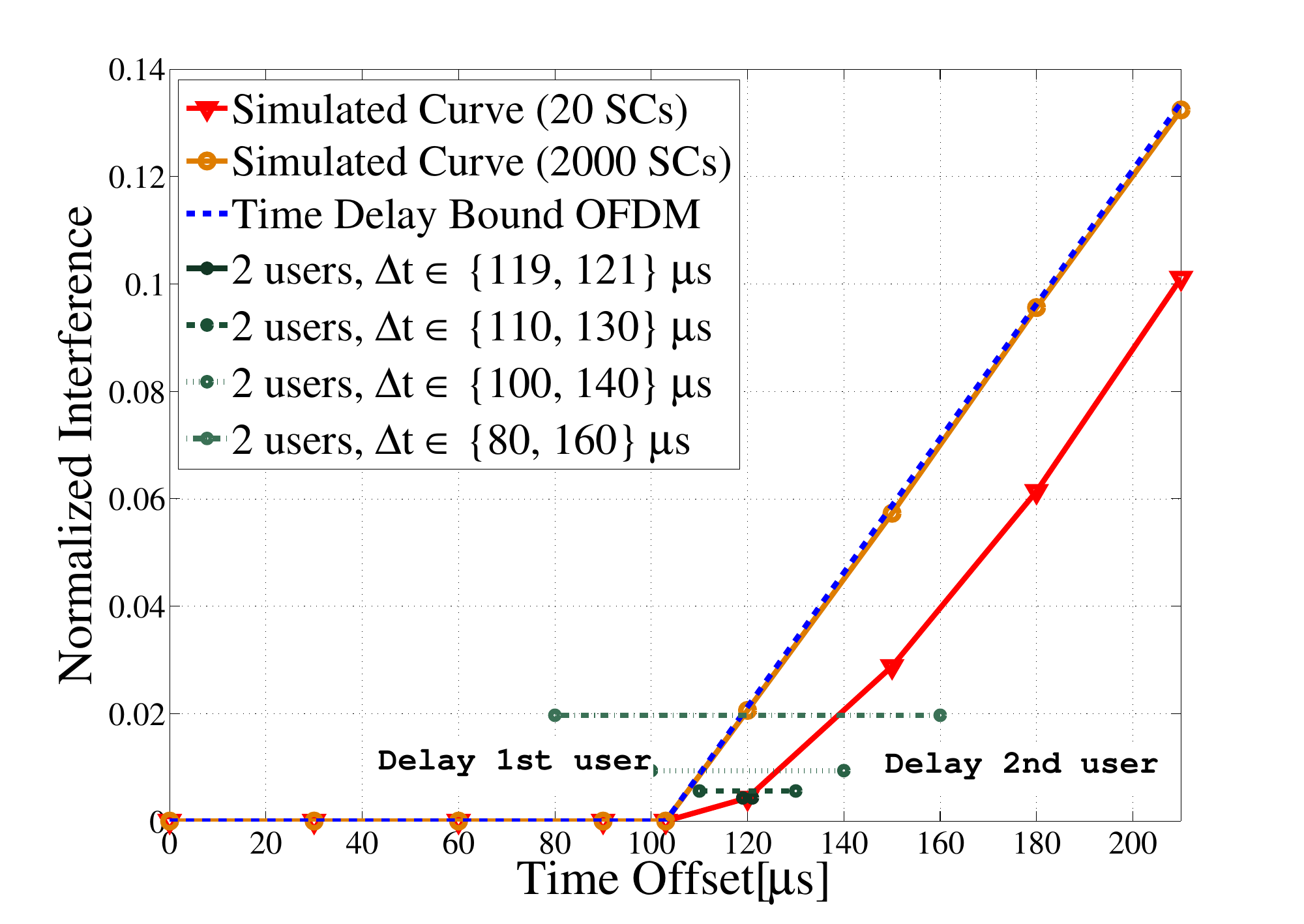}\caption{Simulated
interference over time offsets using different numbers of subcarriers,
compared to theoretical bound. }%
\label{fig:IntVsTimeOff_OFDM}%
\end{figure}

Let us also consider time-invariant channels, i.e., (distributional) spreading
functions of the form $\boldsymbol{\Sigma}(\mu)=h(\mu_{1})\delta(\mu_{2})$.
Here, $h$ is called the channel impulse response. If the system exhibits a
timing offset $\nu_{1}$ with $[\nu_{1}+{\tau_{d}}]_{\text{cp}}=0$ only, the
time dependency in the cross ambiguity function cancels, thus
\begin{equation}
s_{m}((\nu_{1},0))=e^{-i2\pi\nu_{1}m_{2}/T_{u}}{\mathbf{A}}_{g\gamma}((\nu
_{1},0))\newline=e^{-i2\pi\hat{\nu}_{1}m_{2}/\epsilon}{\mathbf{A}}_{g\gamma
}((\nu_{1},0))=e^{i[\phi_{0}-2\pi\hat{\nu}_{1}m_{2}/\epsilon]}%
\end{equation}
and only phase rotations occur (normally corrected by channel estimation and
equalization). Contrary to this, time offsets with $[\nu_{1}+{\tau_{d}%
}]_{\text{cp}}\neq0$ cause interference. For frequency offsets, interference
occurs immediately. For the case $[\nu_{1}+{\tau_{d}}]_{\text{cp}}=0$, the
following relation holds:
\begin{equation}
\langle\boldsymbol{\Sigma},s_{m}\rangle=e^{-i2\pi\lbrack\hat{\nu},m]/\epsilon
}\frac{\sin\pi\hat{\nu}_{2}}{\pi\hat{\nu}_{2}}e^{i\phi_{0}}\,\hat{h}%
(m_{2}/T_{u}).
\end{equation}
Obviously, the frequency offset $\hat{\nu}_{2}$ (the time offset $\hat{\nu}%
_{1}$) induces a rotating phase over the time slots $l$ (over frequency slots
$k$) as one would expect.

Finally, we need the Bessel bound (by contrast to transmit pulse) for the
receive pulse $g$, which is $B_{g}=1$ (see Example 2), and the energy constant, 
which is $E_{g}=\epsilon$, so that altogether
\begin{equation}
\frac{1}{\left\vert {\mathcal{I}}\right\vert }\mathbb{E}\sum_{m}(I_{m}%
+D_{m})\leq\epsilon-\frac{1}{\left\vert {\mathcal{I}}\right\vert }\sum
_{m\in\mathcal{I}}\frac{\sin^{2}(\pi\nu_{2}^{m}(T_{u}-|[\nu_{1}^{m}%
]_{\text{cp}}|))}{(\pi\nu_{2}^{m}T_{u})^{2}}\chi_{\lbrack-T_{u},T_{u}]}%
([\nu_{1}^{m}]_{\text{cp}}).
\end{equation}

\subsubsection{Evaluation}

First, we consider only frequency offsets with no additional delay in time.
Figure \ref{fig:IntVsFrOff_OFDM} shows a comparison of this interference bound
with simulated curves at different numbers of subcarriers. It can be observed
that with an increasing number of subcarriers considered, the interference
curve gets closer to the theoretical bound. In case of 200, or even more
20000, subcarriers, the simulations match the bound almost perfectly.

The described curves are based on a single frequency offset only, which is the
same for all subcarriers. However, Figure \ref{fig:IntVsFrOff_OFDM}
additionally illustrates the behavior in case of multiple different offsets.
For the sake of illustration, we demonstrate the case of two offsets here,
where each offset applies to an equal share of the available subcarriers. It
can be observed that with decreasing difference in the offsets, the resulting
interference level gets closer to value of the bound at the corresponding
average of the offsets.

Let us now consider the case of time delays. Figure
\ref{fig:IntVsTimeOff_OFDM} compares the interference caused by the
asynchronous mode of operation to that predicted by the theoretical bound.
Again, it can be observed that the numerical results converge to the bound
when increasing the number of subcarriers. Note that the CP length is 103ms;
an smaller offset does not produce any interference (however, negative delays
do, which is not depicted here).

\subsection{Spline-based modulation}

\begin{figure}[ptb]
\centering\includegraphics[width=0.9\linewidth]{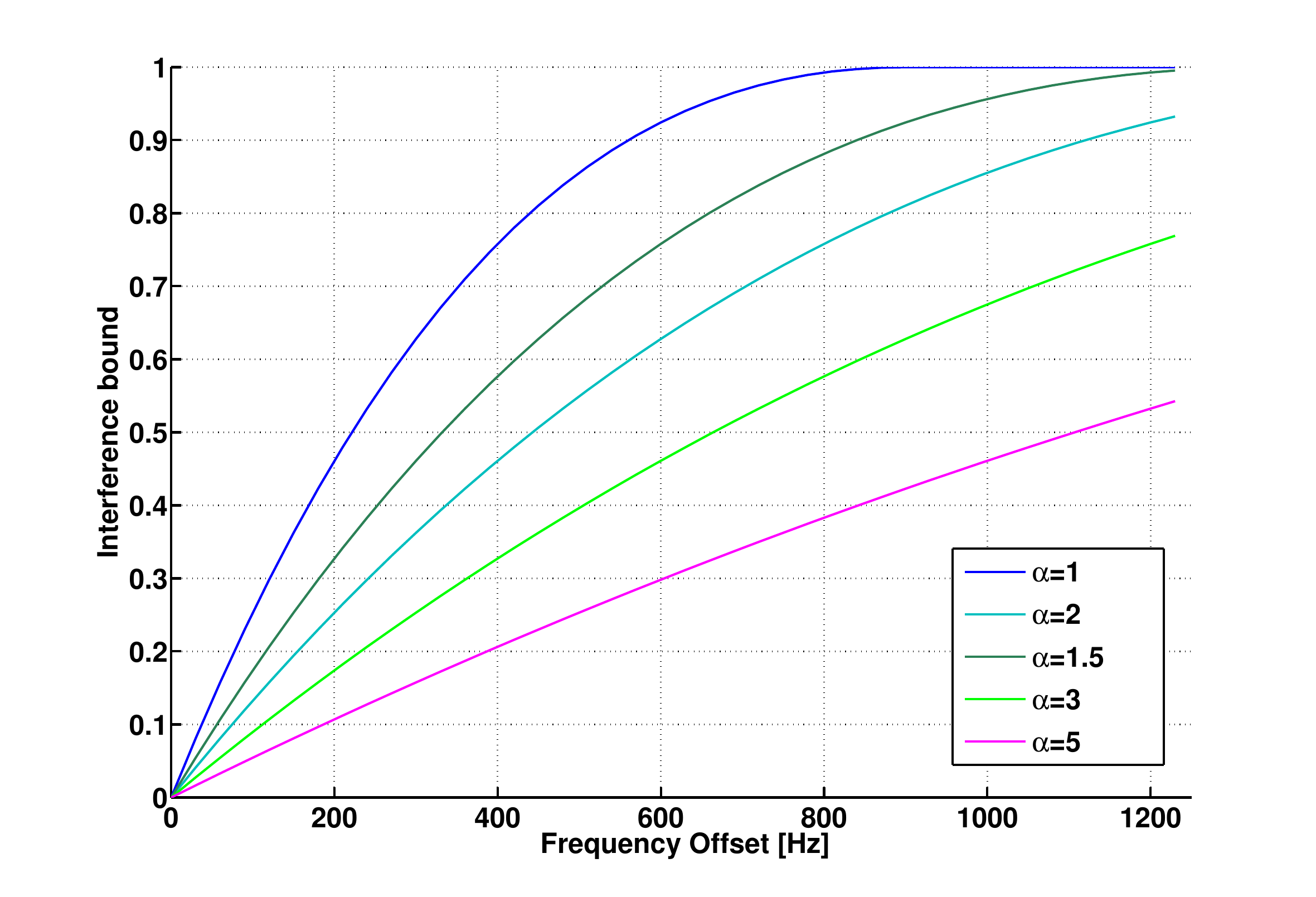}
\caption{Interference bound for spline waveform based over frequency offset
for increasing values of alpha. Note that $B_{\gamma}=1$ is assumed in all
curves.}%
\label{fig:CompAlphaVsFrOff_Spline}%
\end{figure}

Let us now investigate the influence of time and frequency offsets on the
spline-based waveform, as carried out for OFDM in Section
\ref{sec:timeandfreqoffsets}. In order to obtain a bound on the influence of
time or frequency offsets, we need to evaluate the cross-ambiguity function
$A_{g\gamma}$. For this, the dual pulse has to be taken into account. However,
since this is not desirable, simple estimates of this function are needed. For
this purpose, consider the following bound. Define $\gamma_{\mu}%
:=\boldsymbol{S}_{\mu}\gamma$ and $\gamma_{\nu}:=\boldsymbol{S}_{\nu}\gamma$, 
and using $\langle g,\gamma\rangle=1$ and $\lVert g\rVert_{2}=E_{g}$: %
\begin{align}
|A_{g\gamma}(\mu)|  &  =|\langle g,\gamma_{\mu}-\gamma+\gamma\rangle|\\
&  =|1+\langle g,\gamma_{\mu}-\gamma\rangle|=|1+\langle g,\gamma_{\mu}%
-\gamma_{\nu}+\gamma_{\nu}-\gamma\rangle|\\
&  \geq1-|\langle g,\gamma_{\mu}-\gamma_{\nu}\rangle|-|\langle g,\gamma_{\nu
}-\gamma\rangle|\\
&  \geq1-E_{g}\lVert\gamma_{\mu}-\gamma_{\nu}\rVert_{2}-E_{g}\lVert\gamma
_{\nu}-\gamma\rVert_{2},%
\end{align}
where the RHS constitutes a similarity measure for $\gamma$. Let us take
$\mu=(\Delta t,\Delta\omega)$ and $\nu=(0,\Delta\omega)$ and it follows:

\begin{theorem}
Suppose $\lVert\gamma_{m}\rVert_{2}^{2}=1$ (without loss of
generality)$,\lVert g_{m}\rVert_{2}^{2}=E_{g}$ such that $\langle
\boldsymbol{g}_{m},\gamma_{m}\rangle=1$ (perfect reconstruction in noiseless
case). Then
\begin{equation}
|A_{g\gamma}(\mu)|\geq1-\lVert\gamma-\gamma(\cdot-\Delta t)\rVert_{2}%
-\lVert\hat{\gamma}-\hat{\gamma}(\cdot-\Delta\omega)\rVert_{2}, %
\end{equation}
where $\hat{\gamma}$ denotes the Fourier transform of $\gamma$.
\end{theorem}

From this, we get by standard analysis the following approximation for
frequency offsets:%
\begin{equation}
A_{g\gamma}((0,\Delta\omega))\geq1-\frac{\sqrt{3}\Delta\omega T}{2\pi\alpha
}\sqrt{\left(  1-\frac{\Delta\omega T}{2\pi\alpha}\right)  }. %
\end{equation}
Using this approximation we can calculate the interference part in
(\ref{eq:interference_bound}). Figure \ref{fig:CompAlphaVsFrOff_Spline}
illustrates the behavior of the spline waveform with different pulse widths
$\alpha$. It can be observed, that the interference decreases with increasing
pulse width. Similar to frequency offsets, we can derive:%

\begin{equation}
A_{g\gamma}((\Delta t,0))\geq1-\frac{2\pi\alpha\Delta t}{\sqrt{20}T}.%
\end{equation}

\subsubsection{Evaluation}

\begin{figure}[t]
\centering\includegraphics[width=0.9\linewidth]{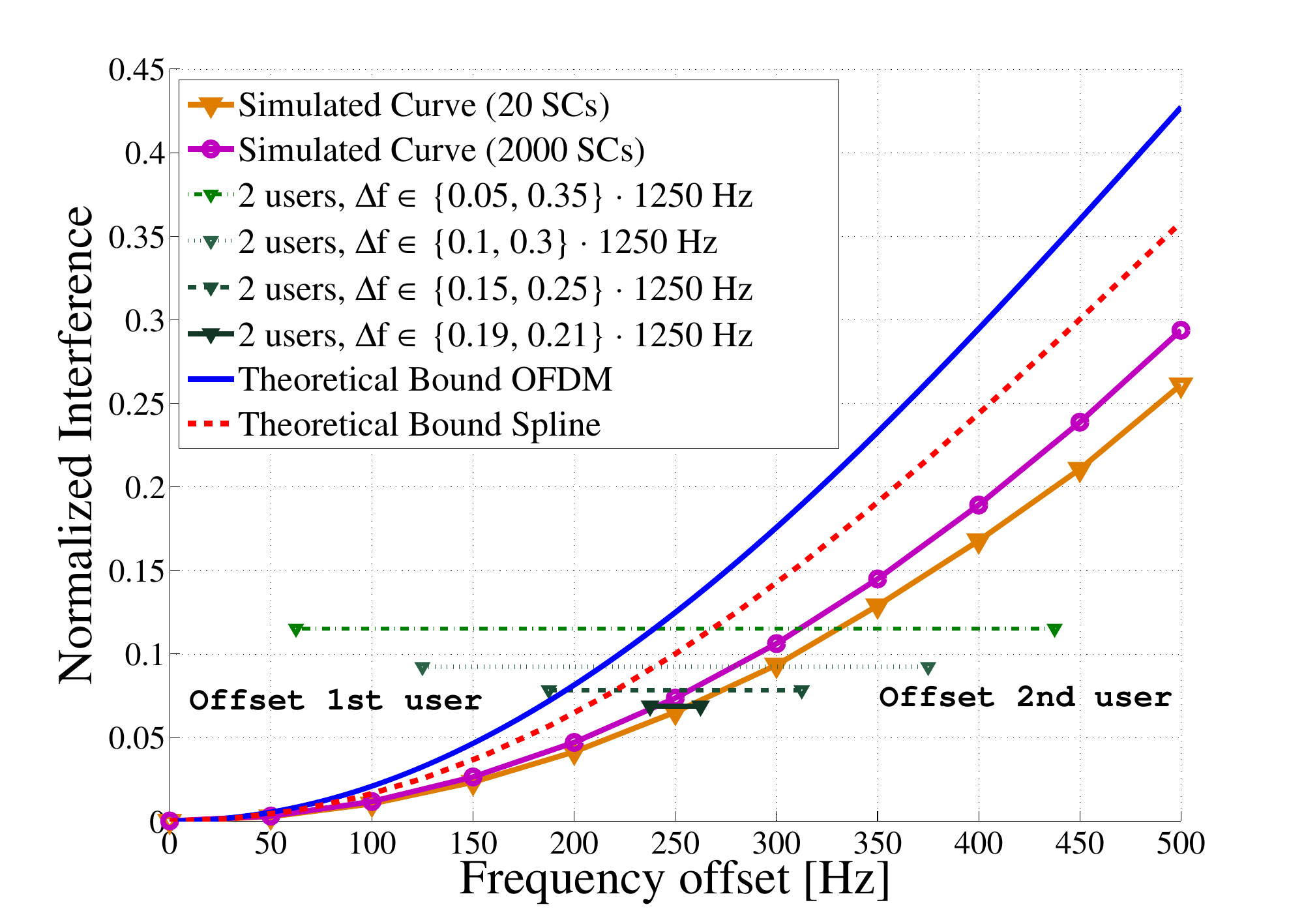}
\caption{Simulated interference and interference bound vs. frequency offsets
for the spline waveform. Numerical results for different numbers of
subcarriers are shown. To foster an easy comparison, the interference bound
for OFDM is also depicted.}%
\label{fig:IntSplineFreqOffs}%
\end{figure}

\begin{figure}[t]
\centering\includegraphics[width=0.9\linewidth]{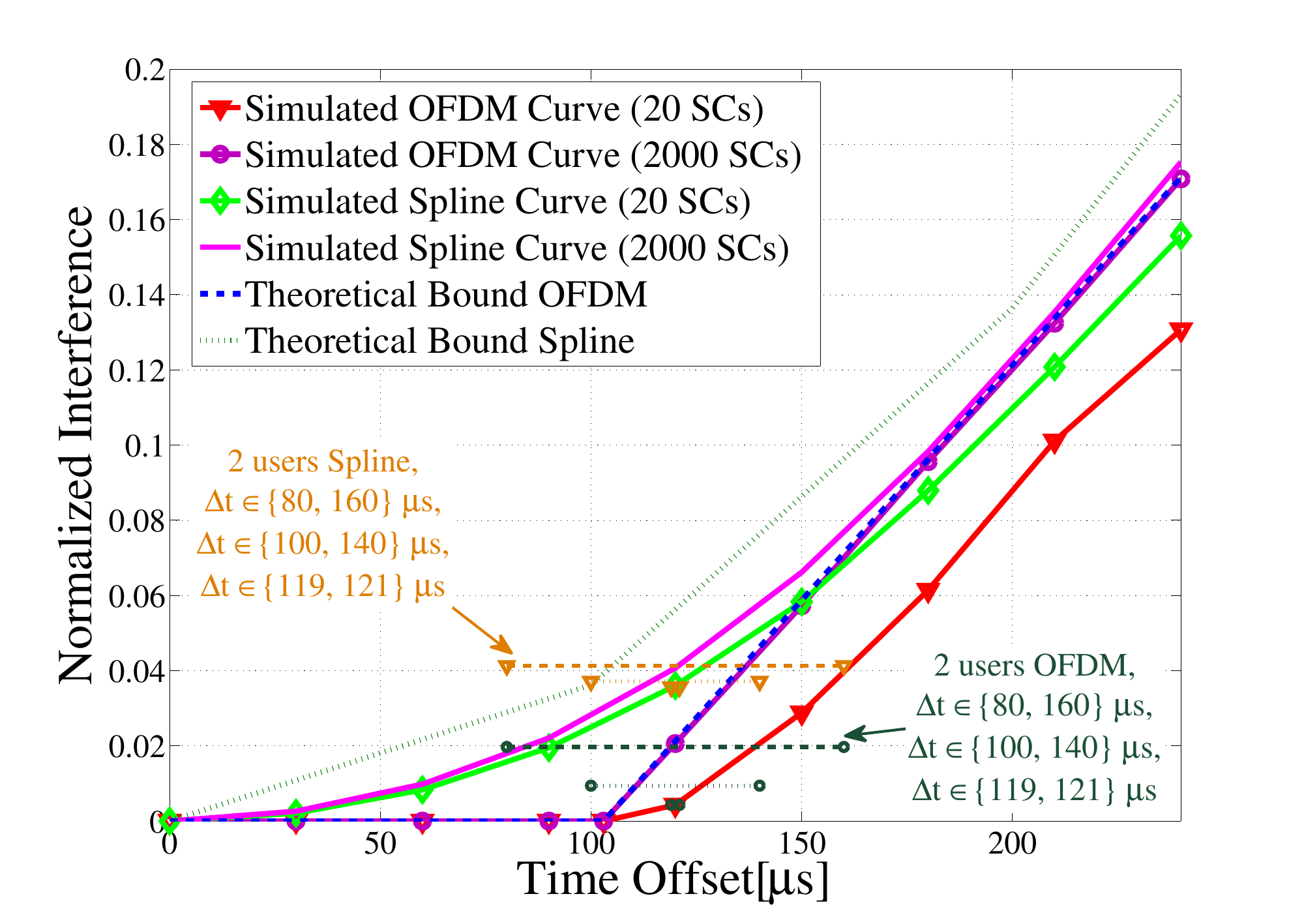}
\caption{Simulated interference and interference bound vs. time offset for the
spline waveform. Numerical results for different numbers of subcarriers are
shown. To foster an easy comparison, the interference bound for OFDM is also
depicted.}%
\label{fig:InteSplineTimeOffs}%
\end{figure}

In Figure \ref{fig:IntSplineFreqOffs}, we show the simulated interference of
the spline waveform together with the corresponding bound based on a numerical
computation of $A_{g\gamma}$ in (\ref{eq:interference_bound}) for frequency
offsets. In addition, the figure again shows also the interference bound for
OFDM. While, the results for the spline waveform indicate lower interference
than in the OFDM case, the bound appears to be less tight, even with large
numbers of subcarriers.

\begin{figure}[h]
\centering\includegraphics[width=0.9\linewidth]{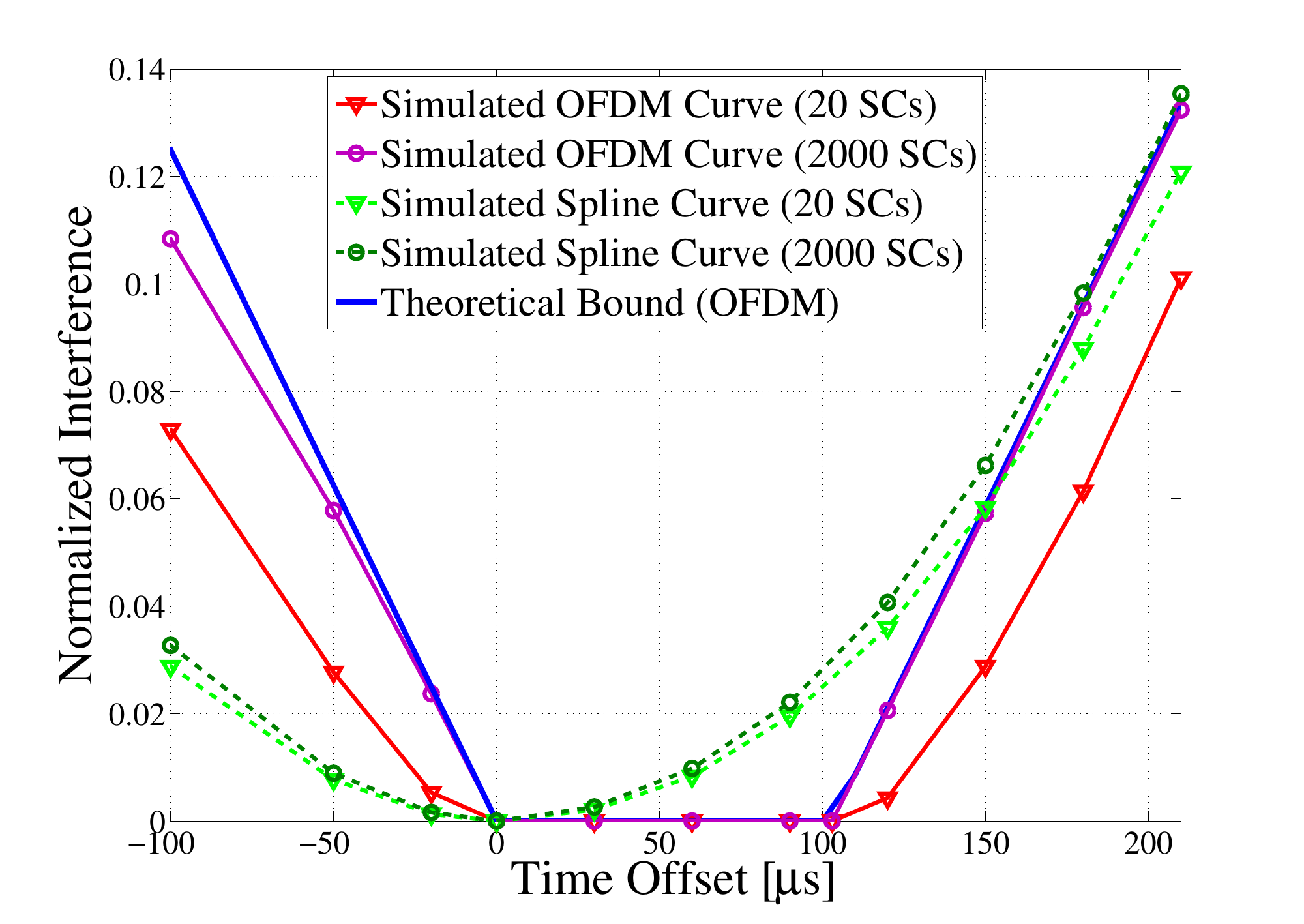}
\caption{Simulated interference and interference bound vs. negative and
positive time offsets for OFDM and spline waveform. Numerical results for
different numbers of subcarriers are shown. To foster an easy comparison, the
interference bound for OFDM is also depicted.}%
\label{fig:PosAndNegTimeOffs}%
\end{figure}

Let us now consider the case of time offsets. Figure
\ref{fig:InteSplineTimeOffs} depicts the interference bound, again based on a
numerical calculation of $A_{g\gamma}$ in (\ref{eq:interference_bound}) and
the simulated interference vs. a time offset for the spline waveform. To allow
an easy comparison, the bound for OFDM is also shown. It should be noted that
although the results do not outperform OFDM for the positive delays considered
in Figure \ref{fig:InteSplineTimeOffs}, the behavior is different for negative
delays. In Figure \ref{fig:PosAndNegTimeOffs}, we show the influence of
negative delays, where the benefits of the spline waveform become obvious.

\section{Practical Implementation}

\label{sec:implementation}

In the following, let $T_{s}$ denote the sampling period, which is equal to
$1/f_{s}$, with $f_{s}$ being the sampling frequency. In the following
discrete model, we let all time indices be multiples of $T_{s}$ and frequency
indices be multiples of $F$. Furthermore, we use $N$ to denote the discrete
counterpart of the symbol duration and submit $K$ symbols. Let
$N_{\mathrm{FFT}}$ be the FFT-length in D-PRACH. Note, for some numerical
reasons $N_{\mathrm{FFT}}$ must divide $NK$. In order to be compliant with 4G, 
we set $TF=1.25$. Note that efficient implementations are available in
literature \cite{vangelista2001efficient}.

\subsection{Transmitter}

For the pulse shaped PRACH, additional processing is needed, compared to
standard OFDM. In contrast to standard processing, we process more than one
symbol interval, even if we use only one symbol to carry the preamble. A pulse
$g$ is used to shape the spectrum of the preamble signal (which is constructed
from a Zadoff-Chu (ZC) sequence \cite{chu1972polyphase}), e.g., to allow the
use of PRACH guard bands with acceptable interference. Let $P$ be the length
of pulse $g$. We extend the output signal $s[n]$ after the inverse FFT (IFFT)
stage by repeating it and taking modulo $P$ to get the same length as the
pulse $g$. Given $K$ symbols, each symbol $s_{k}[n]$ is pointwise multiplied
by the shifted pulse $g$ and superimposed by overlap and add, such that we get
the baseband pulse shaped D-PRACH transmit signal as:%
\begin{equation}
s[n]=\sum_{k=0}^{K-1}s_{k}[n]g[n-kN].
\end{equation}
In greater detail, this can be written as
\begin{equation}
s[n]=\beta\sum_{k=0}^{K-1}\sum_{l\in\mathcal{F}_{1}\cap\mathcal{F}_{2}}%
\tilde{X}_{k,l}\;g[n-kN]e^{j\frac{2\pi nl}{N_{\mathrm{FFT}}}},
\end{equation}
where $\tilde{X}_{k,l}$ is the Fourier transformed preamble signal (which is
constructed from Zadoff-Chu (ZC) root sequences \cite{chu1972polyphase} in
LTE-A) occupying subcarriers $l\in\mathcal{F}_{1}$ of length $N_{\mathrm{ZC}}$
and the actual data occupying subcarriers $l\in\mathcal{F}_{2}$ at the $k$-th
symbol, and $\beta$ is an amplitude scaling factor for normalizing the
transmit power.

\subsection{Receiver}

Assuming an AWGN channel and one user transmitting its preamble signal on the
PRACH, the base station obtains the superposition of data bearing signals,
preamble signal, and noise as%
\begin{equation}
r[n]=s_{\mathrm{PU}}[n]+s_{\mathrm{PR}}[n]+n_{0}[n], \label{eq:receive}%
\end{equation}
where $s_{\mathrm{PU}}$ is the PUSCH data transmit signals, $s_{\mathrm{PR}}$
is the PRACH preamble transmit signal, and $n_{0}$ is Gaussian noise.

The only difference of our BFDM receiver to the standard OFDM PRACH receiver
is the processing before the FFT. While in standard PRACH processing the CP
has to be removed from the received signal $r_{\mathrm{PR}}[n]$, in our BFDM
receiver, an operation to invert the (transmitter side) pulse shaping is
carried out. To be more precise, the $K$ symbols of the received signal
$r_{\mathrm{PR}}[n]$ are pointwise multiplied by the shifted bi-orthogonal
pulse $\gamma$, such that we have
\begin{equation}
r_{k}^{\gamma}[n]=r_{k}[n]\gamma^{\ast}[n-kN].
\end{equation}
Subsequently, some kind of prealiasing operation is applied to each windowed
$r_{k}^{\gamma}[n]$, i.e.,
\begin{equation}
\tilde{r}_{k}^{\gamma}[n]=\sum_{l=0}^{P/N_{\mathrm{FFT}}-1}r_{k}^{\gamma
}[n-lN_{\mathrm{FFT}}],
\end{equation}
such that we obtain the Fourier transformed preamble sequence at the $k$-th
symbol and $l$-th subcarrier after the FFT operation
\begin{equation}
\tilde{Y}_{k,l}=\sum_{n=0}^{N_{\mathrm{FFT}}-1}\tilde{r}_{k}^{\gamma
}[n]e^{-\frac{j2\pi nl}{N_{\mathrm{FFT}}}}.
\end{equation}
Although we do not employ a CP as in standard PRACH, the time-frequency
product of $TF=1.25$ allows the signal to have guard regions in time and
frequency as well. This time-frequency guard regions and the overlapping of
the pulses evoke the received signal to be cyclostationary \cite{Bol01}, which
gives the same benefit as the cyclostationarity made by CP. Furthermore, it is
also shown in \cite{Bol01}, that the bi-orthogonality condition of the pulses
is sufficient for the cyclostationarity and makes it possible to estimate the
symbol timing offset from its correlation function.

\subsection{User Detection}

\label{sec:userDetect}

\subsubsection{Preamble generation}

The preamble is constructed from a ZC sequence as
\begin{equation}
x_{u}[m]=\exp\left\{  -j\frac{\pi um(m+1)}{N_{\mathrm{ZC}}}\right\}  ,0\leq
m\leq N_{\mathrm{ZC}}-1,
\end{equation}
where $u$ is the root index and $N_{\mathrm{ZC}}$ is the length of the
preamble sequence, which is fixed for all users. Here, we consider the case of
contention based RACH, where every user wanting to send a preamble chooses a
signature randomly from the set of available signatures $\mathcal{S}%
=\{1,...,64-N_{\mathrm{cf}}\}$, with $N_{\mathrm{cf}}$ being a given number of
reserved signatures for contention free RACH. Every element of $\mathcal{S}$
is assigned to index $(u,v)$, such that the preamble for each user is obtained
by cyclic shifting the $u$-th Zadoff-Chu sequence according to $x_{u,v}%
[m]=x_{u}[(m+v\,N_{\mathrm{CS}})\,\mathrm{mod}\,N_{\mathrm{ZC}}]$, where
$v=1,...,\left\lfloor \frac{N_{\mathrm{ZC}}}{N_{\mathrm{CS}}}\right\rfloor $
is the cyclic shift index and $N_{\mathrm{CS}}$ is the cyclic shift size.
Since only $V=\left\lfloor \frac{N_{\mathrm{ZC}}}{N_{\mathrm{CS}}%
}\right\rfloor $ preambles can be generated from the root $u$, the assignment
from $\mathcal{S}$ to $(u,v)$ depends on $N_{\mathrm{CS}}$ and on the size of
set $\mathcal{S}$.

\subsubsection{Signature detection}

Given the received signal (\ref{eq:receive}), the PRACH receiver observes the
fraction $y$ that lies in the PRACH region to obtain the preamble. The
receiver stores all available Zadoff-Chu roots as a reference. These root
sequences are transformed to frequency domain and each of them is multiplied
with the received preamble. As discussed in Section \ref{sec:Gabor}, it
approximately holds, as in OFDM,
\begin{equation}
Z_{u}[w]=Y[w]X_{u}^{\ast}[w],
\end{equation}
where $Y[w]$ is the received preamble and $X_{u}[w]$ is the $u$-th ZC sequence
in frequency domain respectively. Using the convolution property of the
Fourier transform it is easy to show that $Z_{u}[w]$ is equal to the inverse
Fourier transform of any cross correlation function $z_{u}[d]$ at lag $d$.
Because the preamble is constructed by cyclic shifting the Zadoff-Chu
sequence, ideally we can detect the signature by observing a peak from the
power delay profile, given by
\begin{equation}
\left\vert z_{u}[d]\right\vert ^{2}=\left\vert \sum_{n=0}^{N_{\mathrm{ZC}}%
-1}y[n+d]x^{\ast}[n]\right\vert ^{2}. \label{eq:pdp}%
\end{equation}
Let $N_{\mathrm{root}}=\left\lfloor \frac{64-N_{\mathrm{cf}}}{V}\right\rfloor
$ be the number of roots that we require to generate $64-N_{\mathrm{cf}}$
preambles. Then, the signature $S_{i}$ and the delay $d_{i}$ of user $i$ are
obtained by $S_{i}=Vu+\left\lfloor \frac{\tau_{l}}{N_{\mathrm{CS}}%
}\right\rfloor ,\;(0\leq u\leq N_{\mathrm{root}})$ and $d_{i}=(\tau
_{l}\;\mathrm{mod}\;N_{\mathrm{CS}})\times\frac{N_{\mathrm{FFT}}%
}{N_{\mathrm{ZC}}}\;T_{s}$, respectively, where $\tau_{l}$ is the location of
the largest peak in (\ref{eq:pdp}).

\subsection{Channel Estimation}

\label{sec:chEst}

The question remains how to obtain an estimation for the channel also on the
new D-PRACH subcarriers. Due to our system setup, we assume that the received
preamble signal can be written as
\begin{equation}
y=\underbrace{D\cdot W}_{\Phi}\cdot h+e. \label{eq:system_ch_est}%
\end{equation}
Thereby, the term $e$ accounts for all interference and noise, $D$ is a
diagonal matrix constructed from the coefficients of the Fourier transformed
preamble and $W=F(\mathcal{I}_{p},\mathcal{I}_{h})$ is a sub-matrix of the
$\mathbb{C}^{M\times M}$ DFT-matrix $F$. The set $\mathcal{I}_{h}%
:=\{1,\ldots,n_{h}\}$ contains the indices of the first $n_{h}$ columns, and
$\mathcal{I}_{p}=\{i_{1},\ldots,i_{N_{\mathrm{ZC}}}\}$ contains the indices of
the central $N_{ZC}$ rows of $F$. Furthermore, $M$ is the length of the
subframe without CP and guard interval, and we assume a maximum length $n_{h}$
of the channel $h$.

For simplicity, we consider simple least-squares channel estimation, i.e., we
have to solve the estimation (normal) equation $\Phi^{H}\Phi\hat{h}=\Phi^{H}%
y$. To handle cases where $\Phi$ is ill-conditioned, we use Tikhonov
regularization. This popular method replaces the general problem of $\min
_{x}\lVert Ax-y\rVert^{2}$ by $\min_{x}\lVert Ax-y\rVert^{2}+\lVert\Gamma
x\rVert^{2}$, with the regularization matrix $\Gamma$. In particular, for our
model in (\ref{eq:system_ch_est})
\begin{equation}
\hat{h}=(\Phi^{H}\Phi+\Gamma^{H}\Gamma)^{-1}\Phi^{H}y
\end{equation}
is used in place of the pseudo-inverse, where $\Gamma$ has to be adapted to
the statistical properties of $e$. We choose $\Gamma$ to be a multiple of the
identity matrix.

The idea behind the estimation approach is, that the estimated channel is also
valid for subcarriers that are adjacent to the region that we actually
estimate the channel for. Numerical experiments indicate that the MSE is
smaller then $10^{-4}$ for up to 200 subcarriers outside of $\mathcal{I}_{p}$.

\section{Performance Evaluation}

\label{sec:simulations}

In this section we verify, using numerical experiments, that using the PRACH
guard bands to carry messages is indeed practicable. We compare the standard
(LTE) PRACH implementation to our proposed spline pulse shaped PRACH.

\subsection{Simulation Setup}

The simulation parameters, chosen according to LTE specifications, are
provided in Table \ref{Tab:spec}. For the computation of $\gamma$ we use the
LTFAT toolbox, which provides an efficient implementation of the $S^{-1}$-trick
\cite{Sondergaard12}. Due to the properties of the pulses, and to fit the
strict LTE frequency specification, we allow a small spillover effect from
PRACH to PUSCH in time. Due to the PRACH pulse length of 4\thinspace ms, as
depicted in Figure \ref{fig:region}, we simulate the PUSCH over this time
interval. Furthermore we use the maximal available LTE bandwidth of
20\thinspace MHz.

\begin{table}[ptb]
\caption{System Specification}%
\label{Tab:spec}
\centering
\begin{tabular}
[c]{|l|l|l|l|}\hline
& PUSCH & standard & pulse shaped\\
&  & PRACH & PRACH\\\hline\hline
Bandwidth & 20\thinspace MHz & 1.08\thinspace MHz & 1.08\thinspace MHz\\\hline
OFDM symbol duration & 0.67\thinspace$\mathrm{\mu s}$ & 800\thinspace
$\mathrm{\mu s}$ & -\\\hline
Subcarrier spacing $F$ & 15\thinspace kHz & 1.25\thinspace kHz &
1.25\thinspace kHz\\\hline
Sampling frequency $f_{s}$ & 30.72\thinspace MHz & 30.72\thinspace MHz &
30.72\thinspace MHz\\\hline
Length of FFT $N_{\mathrm{FFT}}$ & 2048 & 24576 & 24576\\\hline
Number of subcarrier $L$ & 1200 & 839 & 839\\\hline
Cyclic prefix length $T_{\mathrm{CP}}$ & $160\,T_{s}$ 1st & $3168\,T_{s}$ &
0\\
& $144\,T_{s}$ else &  & \\\hline
Guard time $T_{g}$ & 0 & $2976\,T_{s}$ & 0\\\hline
Pulse Length $P$ & - & - & 4\thinspace ms\\\hline
Number of symbols $K$ & 14 & 1 & 1\\\hline
Time-freq. product $TF$ & 1.073 & 1.25 & 1.25\\\hline
\end{tabular}
\end{table}

\begin{figure}[ptb]
\centering
\includegraphics[width=0.9\linewidth]{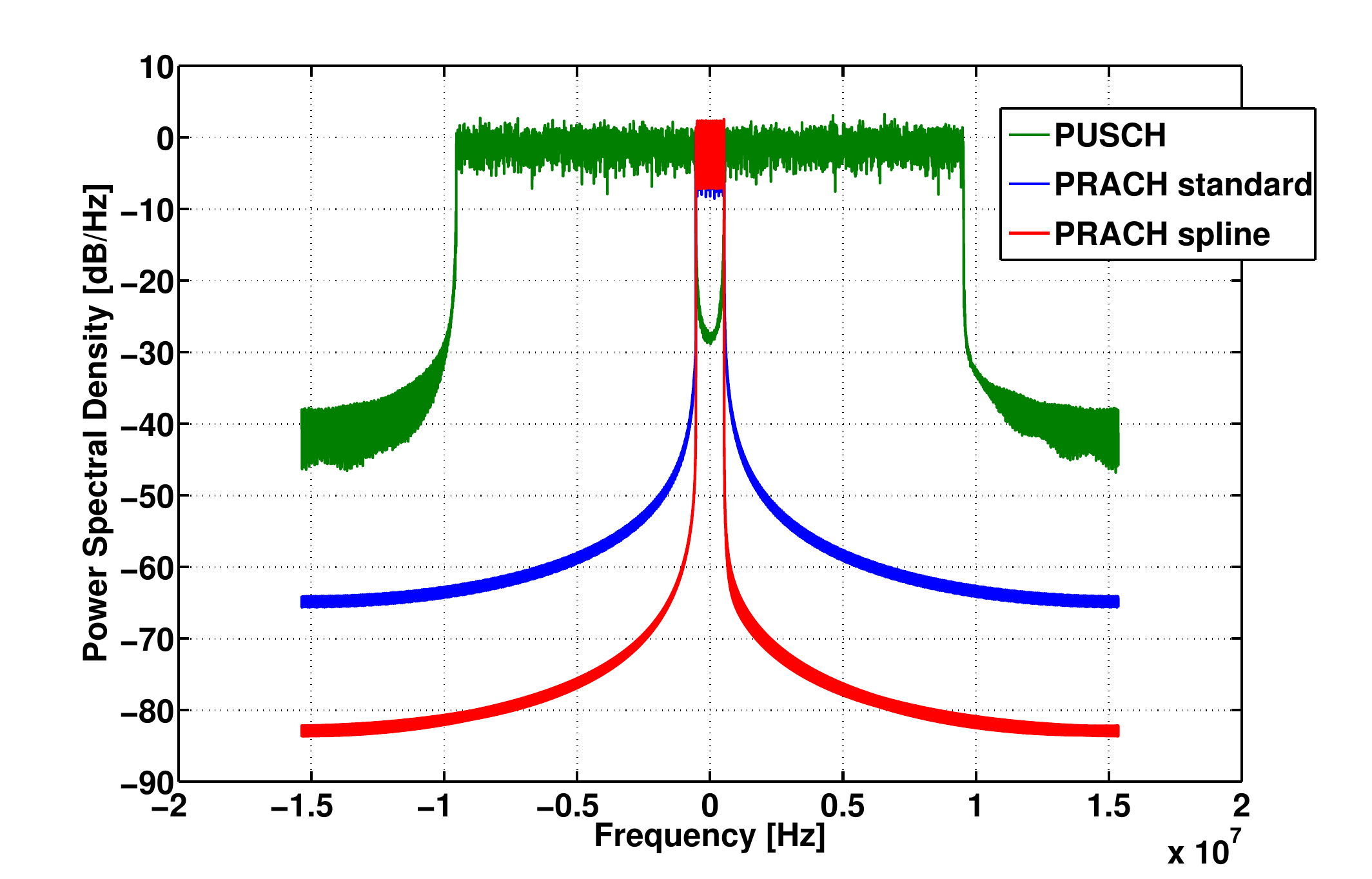} \caption{Power
spectral density. The power of the PRACH is chosen to achieve a similar PSD as
PUSCH.}%
\label{fig:PSD}%
\end{figure}

In the LTE standard, the power of PRACH is variable and is incrementally
increased according to a complicated procedure. To allow a meaningful
comparison without having to implement to complete PRACH procedure, we choose
the power of the PRACH such that approximately the same power spectral density
as in PUSCH is achieved, as depicted in Figure \ref{fig:PSD}. We simulate
multipath channels with a fixed number of three channel taps. Moreover, we
assume a maximum length of $n=300$, which corresponds to a delay spread of
roughly $5\,\mathrm{\mu s}$, and which implies a maximum cell radius of
1.5\thinspace km. For the transmission in PRACH, we use 4-QAM modulation.
Consequently, even in case the PRACH power is lower than in Figure
\ref{fig:PSD}, we still have the opportunity to reduce the modulation to BPSK.

\subsection{Data transmission in PRACH}

\begin{figure}[ptb]
\centering
\includegraphics[width=0.9\linewidth]{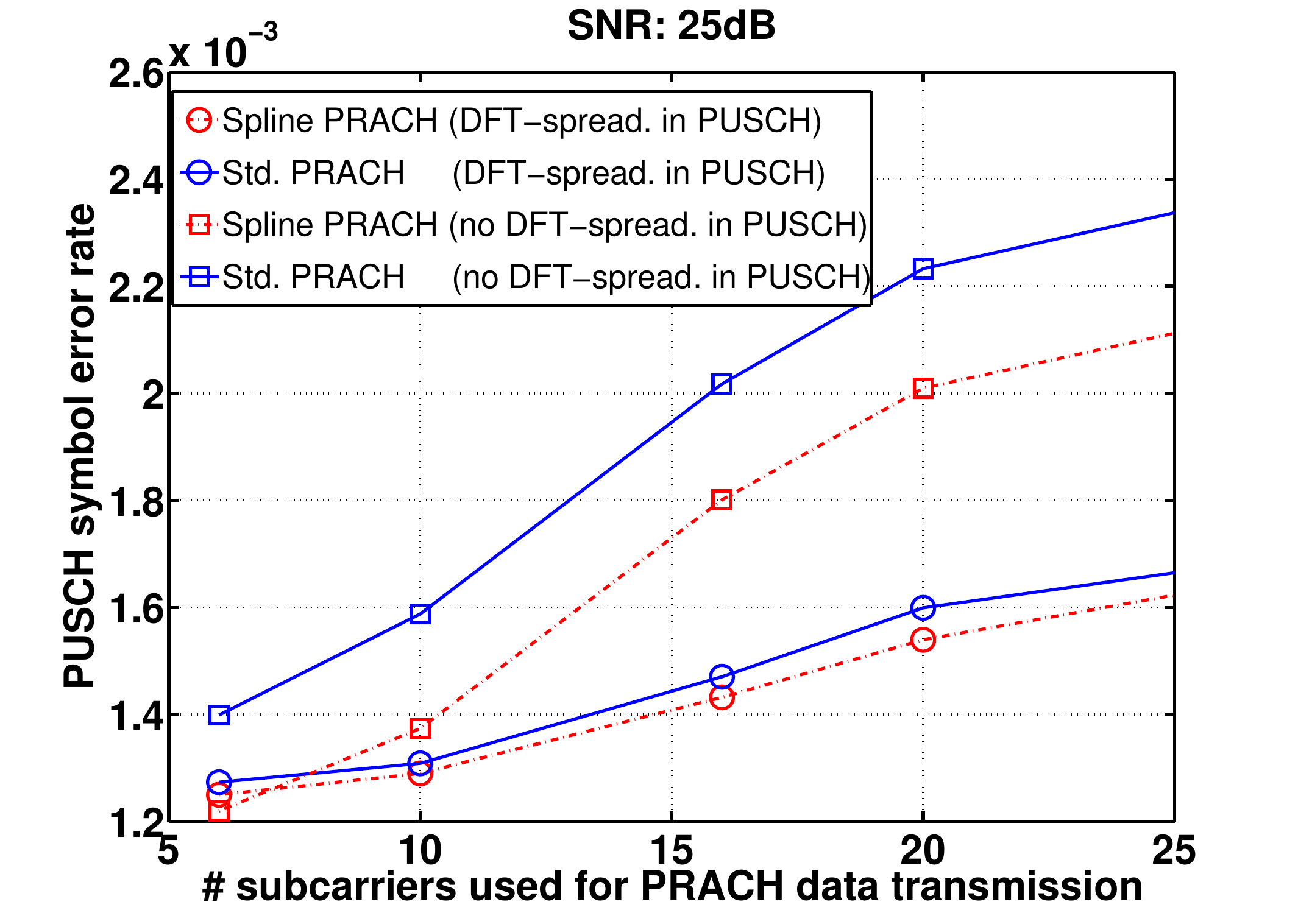}
\caption{Symbol error rate in PUSCH (averaged over all (1200) subcarriers)
plotted over the number of D-PRACH subcarriers. The BFDM-based approach
slightly reduces the symbol error rate. This effect is stronger when no DFT
spreading is performed in PUSCH.}%
\label{fig:PUSER_all}%
\end{figure}

Naturally, using the guard bands for data transmission causes an increased
interference level in PUSCH. In Figure \ref{fig:PUSER_all}, we show the effect
on PUSCH symbol error rate caused by data transmission on a variable number of
D-PRACH subcarriers, given the standard LTE PRACH and the new BFDM-based PRACH approach.

Clearly, the performance of PUSCH does not deteriorate due to the
proposed BFDM-based PRACH. By contrast, irrespective of the actual number of
subcarriers used for data transmission, the BFDM-based approach leads to a
slightly reduced interference level in PUSCH. Due to the strong influence of
the D-PRACH on neighboring subcarriers in PUSCH, this effect is stronger when
no DFT-spreading is used in PUSCH. The reason why larger gains, which could be
expected from Figure \ref{fig:PSD}, cannot be realized is the PUSCH receiver
procedure, which cuts out individual OFDM symbols from the received data.

\subsection{Asynchronous users}

\begin{figure}[ptb]
\centering
\includegraphics[width=0.9\linewidth]{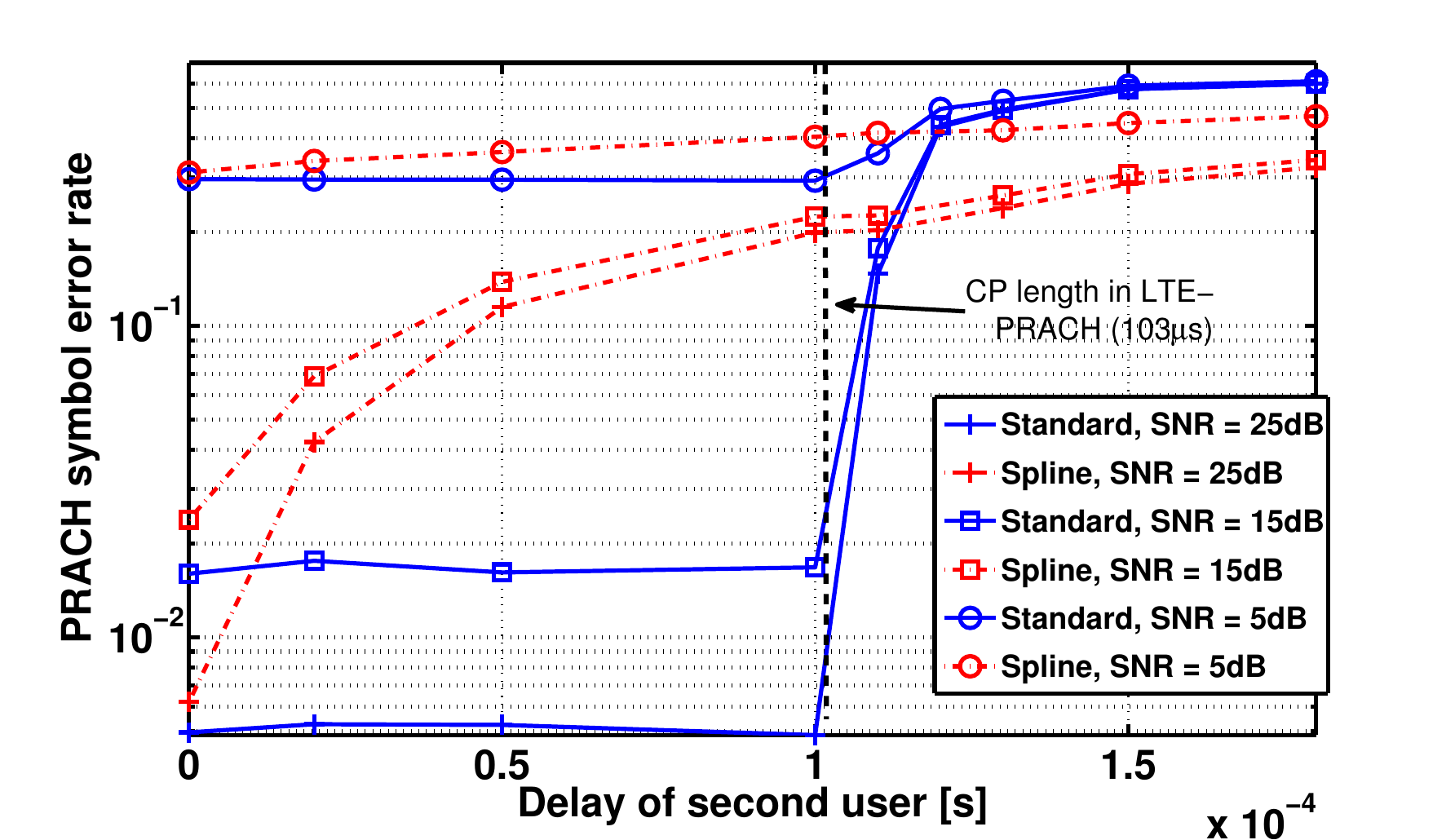}
\caption{Symbol error rate in PRACH (using 4-QAM) averaged over 10 out of 20
data subcarriers vs. the time offset of a second user. The other subcarriers
are used by the second (asynchronous) user. The black line shows the CP length
in LTE PRACH.}%
\label{fig:asynchPRACH}%
\end{figure}

Asynchronous data transmission is a major challenge that comes with MTC and
the IoT. Therefore, we now consider a second, completely asynchronous, user
that transmits data in the PRACH. Thereby we assign half of the subcarriers
available for PRACH data transmission to this second user. However, we still
evaluate only the performance of the original \textquotedblleft user of
interest\textquotedblright\ (and consequently we carry out channel estimation
and decoding only for this user), which is assumed to transmit at the
\textquotedblleft inner\textquotedblright\ subcarriers close to the control
PRACH. Thereby, we compare two waveforms, OFDM and the proposed spline
approach. Figure \ref{fig:asynchPRACH} shows the results.

\begin{figure}
\centering
\includegraphics[width=0.9\linewidth]{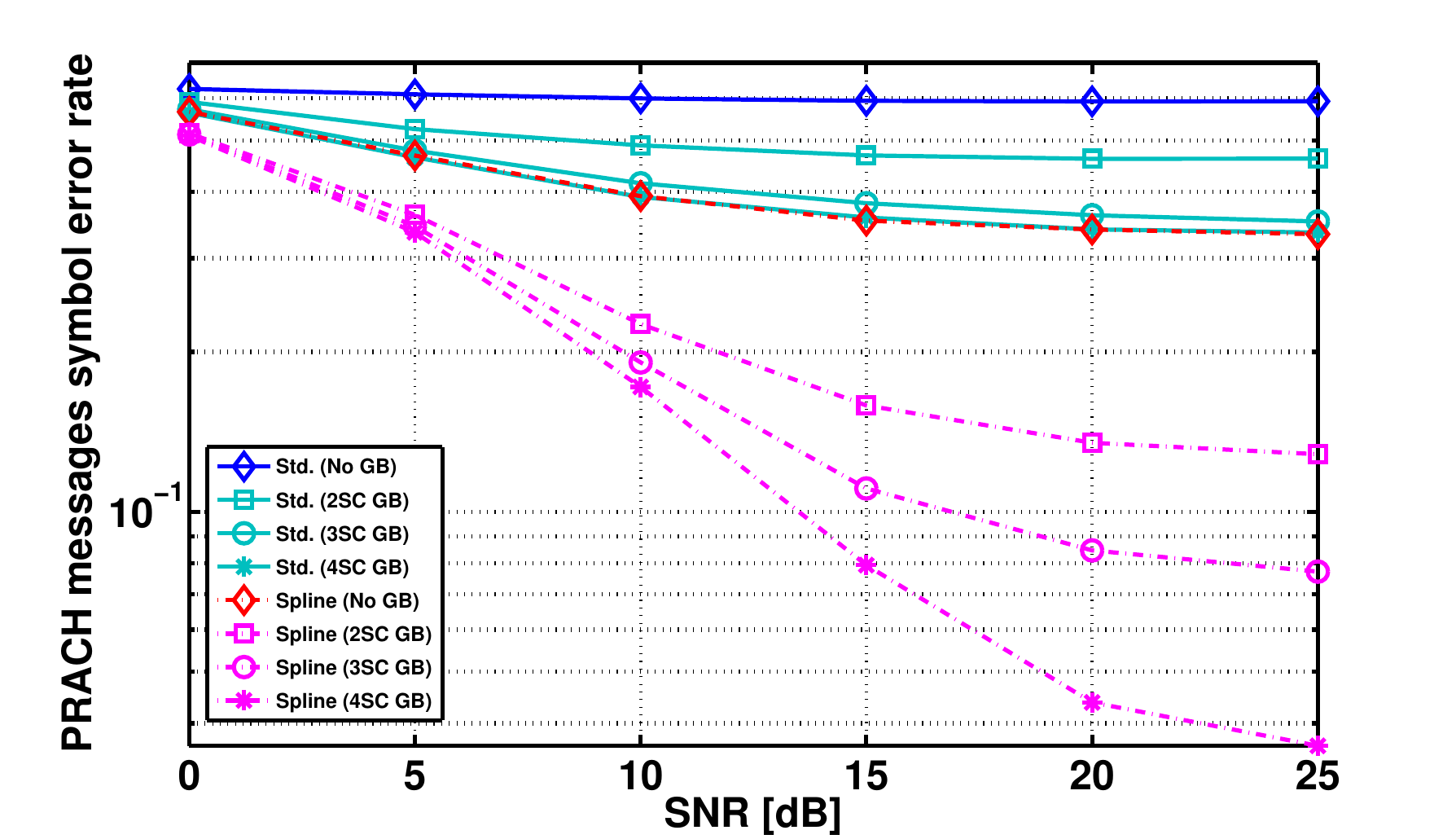} \caption{Symbol error
rate in PRACH (using 4-QAM) over SNR with presence of a second asynchronous
user. Here, the second user has a time offset of $200\,\mu s$. The spline
based approach outperforms OFDM with or without a small number of guard
bands.}%
\label{fig:asynchPRACH_GI}%
\end{figure}

We observe that for completely asynchronous users, i.e., offsets
larger than the CP (in which case OFDM loses its orthogonality property), the
new pulse shaped approach reduced the symbol error rate up to a factor of
almost one half. Nevertheless, the resulting symbol error rate may still seem
excessive. However, as Figure \ref{fig:asynchPRACH_GI} shows, this effect can
be compensated by allowing small guard bands (GB) in between the users. Figure
\ref{fig:asynchPRACH_GI} compares the performance of no GB and GBs of up to 4
subcarriers, which already drastically reduces the symbol error rate.
Interestingly, the spline-based approach without GB achieves roughly the same
performance as OFDM with a GB of 4 subcarriers. In other words, we can save 4
subcarriers using the spline-based PRACH.

Although the symbol error rates depicted in Figure \ref{fig:asynchPRACH} and Figure \ref{fig:asynchPRACH_GI} 
may appear excessive, it should be noted that the delays considered in this evaluations (partially exceeding the cyclic
prefix length) are unusually high. Figure \ref{fig:simulTimeFreqOffsPRACH} gives a different picture, 
focusing on (positive and negative) delays around zero. 

\subsection{Simultaneous time and frequency offsets} \label{sec:Simulation_time_and_freq_offs}

\begin{figure}
\centering
\includegraphics[width=0.9\linewidth]{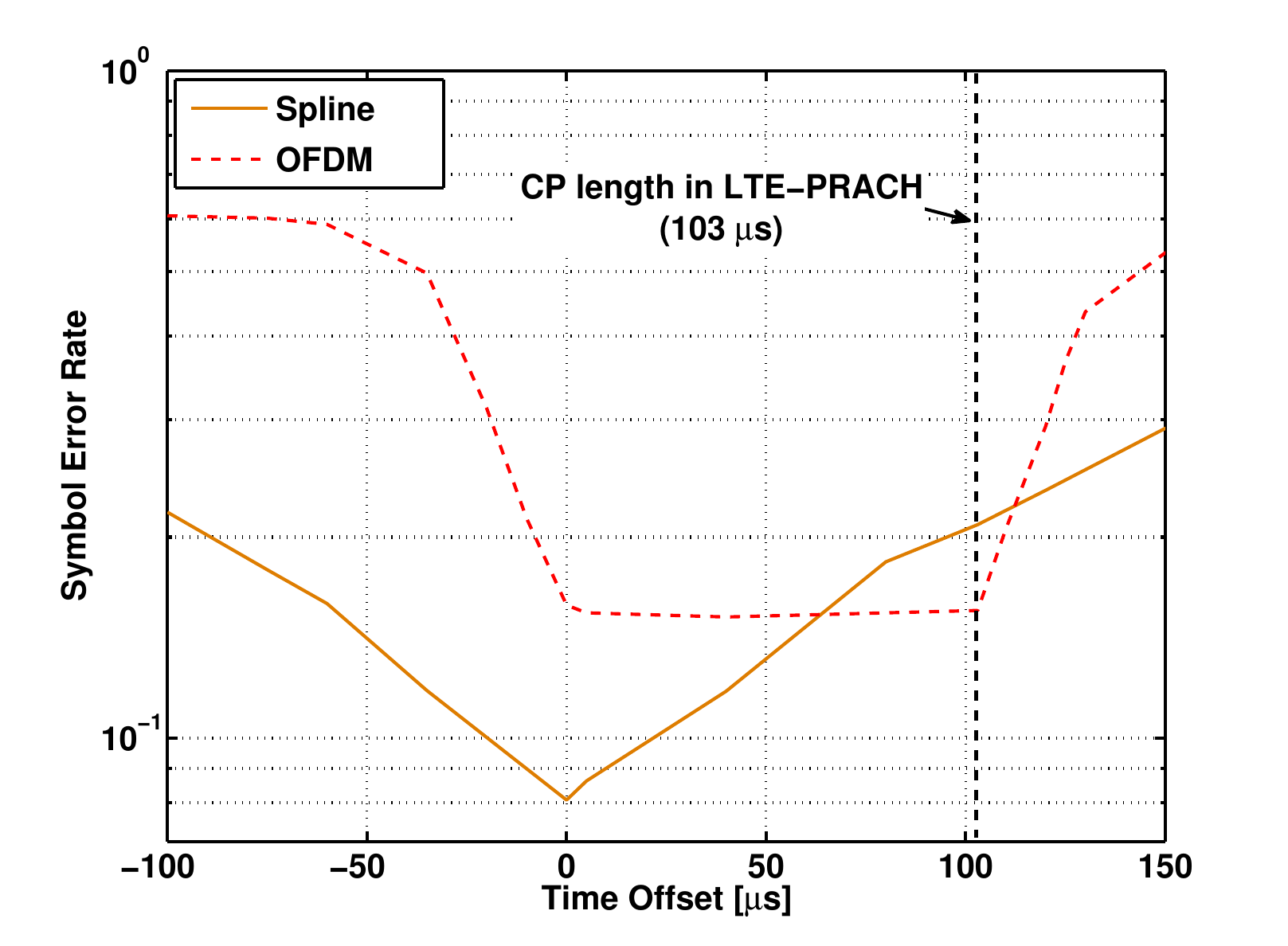}
\caption{Symbol error rate in PRACH (using 4QAM) with perfect channel
knowledge, averaged over 10 out of 20 data subcarriers vs. a varying time
offset of a second user with a frequency offset of 62,5 Hz. The SNR is set to
$25\,\mathrm{dB}$. The black line shows the CP length in LTE PRACH.}%
\label{fig:simulTimeFreqOffsPRACH}%
\end{figure}

Figure \ref{fig:simulTimeFreqOffsPRACH} illustrates the advantages of the
spline-based transmission scheme for both frequency offsets and time delays.
We plot the PRACH symbol error rates over the time offset of a second,
asynchronous user. In addition, a constant small frequency offset of 62,5 Hz
is applied. The SNR is fixed at $25\,\mathrm{dB}$. Moreover, in Figure
\ref{fig:simulTimeFreqOffsPRACH} we assume perfect channel knowledge for the
user of interest. It can be observed that the additional frequency offset has
a detrimental effect on both schemes, however, the performance loss of OFDM is
significantly higher. 

\section{Conclusions}

\label{sec:conclusions}

We proposed and evaluated a novel pulse shaped random access scheme based on
BFDM, which is especially suited in random access scenarios due to very long
symbol lengths. It turns out, that the proposed approach is well suited to
support data transmission within a 5G PRACH. In particular, numerical results
indicate that the BFDM-based approach does not deteriorate PUSCH operations,
in fact, it even leads to a slightly reduced interference in PUSCH when using
(previously unused) guard bands for data transmission, irrespective of the
number of subcarriers used for data transmission. Even more importantly,
completely asynchronous users, with time offsets larger than the CP duration
in standard PRACH, can be far better supported using the BFDM based approach
than using standard OFDM/SCFDMA. The presented results will help to cope with
the upcoming challenges of 5G wireless networks and the IoT, such as sporadic traffic.

\bibliographystyle{IEEEtran}
\bibliography{5GNOW}

\begin{thebibliography}{10}
\providecommand{\url}[1]{#1}
\csname url@samestyle\endcsname
\providecommand{\newblock}{\relax}
\providecommand{\bibinfo}[2]{#2}
\providecommand{\BIBentrySTDinterwordspacing}{\spaceskip=0pt\relax}
\providecommand{\BIBentryALTinterwordstretchfactor}{4}
\providecommand{\BIBentryALTinterwordspacing}{\spaceskip=\fontdimen2\font plus
\BIBentryALTinterwordstretchfactor\fontdimen3\font minus
  \fontdimen4\font\relax}
\providecommand{\BIBforeignlanguage}[2]{{%
\expandafter\ifx\csname l@#1\endcsname\relax
\typeout{** WARNING: IEEEtran.bst: No hyphenation pattern has been}%
\typeout{** loaded for the language `#1'. Using the pattern for}%
\typeout{** the default language instead.}%
\else
\language=\csname l@#1\endcsname
\fi
#2}}
\providecommand{\BIBdecl}{\relax}
\BIBdecl

\bibitem{Kasparick_EW14}
{M. Kasparick, G. Wunder, P. Jung, D. Maryopi}, ``{Bi-Orthogonal Waveforms for
  5G Random Access with Short Message Support},'' in \emph{European Wireless
  Conference (EW'14)}.\hskip 1em plus 0.5em minus 0.4em\relax Barcelona, Spain:
  IEEE Xplore, May 2014.

\bibitem{ComMag5GNOW}
G.~Wunder, P.~Jung, M.~Kasparick, T.~Wild, F.~Schaich, Y.~Chen, S.~ten Brink,
  I.~Gaspar, N.~Michailow, A.~Festag, L.~Mendes, N.~Cassiau, D.~Ktenas,
  M.~Dryjanski, S.~Pietrzyk, B.~Eged, P.~Vago, and F.~Wiedmann, ``{5GNOW:
  Non-Orthogonal, Asynchronous Waveforms for Future Mobile Applications},''
  \emph{IEEE Communications Magazine}, vol.~52, no.~2, pp. 97--105, 2014.

\bibitem{LTE_SIG}
{Alcatel-Lucent, Ericsson, Huawei, Neul, NSN, Sony, TU Dresden, u-blox, Verizon
  Wireless, Vodafone}, \emph{White Paper}.

\bibitem{Schaich_VTC14-Spring}
{F. Schaich, T. Wild, Y. Chen}, ``{Waveform Contenders for 5G - Suitability for
  Short Packet and Low Latency Transmissions},'' in \emph{IEEE 79th Vehicular
  Technology Conference (VTC2014-Spring)}.\hskip 1em plus 0.5em minus
  0.4em\relax Seoul, Korea: IEEE Xplore, May 2014.

\bibitem{Berg_DSD14}
{V. Berg, JB. Doré, and D. Noguet}, ``{A multiuser FBMC Receiver Implementation
  for Asynchronous Frequency Division Multiple Access},'' in \emph{EuroMicro
  DSD2014}.\hskip 1em plus 0.5em minus 0.4em\relax Verona, Italy: IEEE Xplore,
  August 2014.

\bibitem{RACH_ICC14}
G.~Wunder, P.~Jung, and C.~Wang, ``{Compressive Random Access for Post-LTE
  Systems},'' in \emph{IEEE ICC Workshop on Massive Uncoordinated Access
  Protocols}, Sydney, Australia, 2014.

\bibitem{RACH_Metis14}
{Ji, Yalei and Stefanovic, Cedomir, and Bockelmann, Carsten and Dekorsy, Armin,
  and Popovski, Petar}, ``{Characterization of Coded Random Access with
  Compressive Sensing based Multiuser Detection},'' in
  \emph{www.arxiv.com/1404.2119}, 2014.

\bibitem{kozek_98_jsac}
W.~Kozek and A.~Molisch, ``{Nonorthogonal pulseshapes for multicarrier
  communications in doubly dispersive channels},'' \emph{IEEE Journal Sel.
  Areas in Commun.}, vol.~16, no.~8, pp. 1579--1589, 1998.

\bibitem{Sesia:2009}
S.~Sesia, I.~Toufik, and M.~Baker, \emph{{LTE, The UMTS Long Term Evolution:
  From Theory to Practice}}.\hskip 1em plus 0.5em minus 0.4em\relax Wiley
  Publishing, 2009.

\bibitem{ATA_ISWCS14}
F.~Schaich and T.~Wild, ``{Relaxed Synchronization Support of Universal
  Filtered Multi-Carrier including Autonomous Timing Advance},'' in \emph{ISWCS
  Workshop on Advanced Multi-Carrier Techniques for Next Generation Commercial
  and Professional Mobile Systems}, Barcelona, Spain, August 2014.

\bibitem{bello:wssus}
P.~A. Bello, ``{Characterization of randomly time--variant linear channels},''
  \emph{Trans. on Communications}, vol.~11, no.~4, pp. 360--393, 1963.

\bibitem{jung_07_comm}
P.~Jung and G.~Wunder, ``{The WSSUS Pulse Design Problem in Multicarrier
  Transmission},'' \emph{IEEE Trans. on Communications}, 2007.

\bibitem{jung:wcnc08}
\BIBentryALTinterwordspacing
P.~Jung, ``{Pulse Shaping, Localization and the Approximate Eigenstructure of
  LTV Channels},'' \emph{The 2008 IEEE Wireless Communications and Networking
  Conference (WCNC), Las Vegas, USA, Invited Paper}, pp. 1114--1119, 2008.
  [Online]. Available: \url{http://arxiv.org/abs/0912.2828
  http://ieeexplore.ieee.org/search/srchabstract.jsp?tp=\&arnumber=4489232}
\BIBentrySTDinterwordspacing

\bibitem{jung:thesis}
\BIBentryALTinterwordspacing
------, ``{Weyl--Heisenberg Representations in Communication Theory},'' Ph.D.
  dissertation, Technical University Berlin, 2007. [Online]. Available:
  \url{http://opus.kobv.de/tuberlin/volltexte/2007/1619}
\BIBentrySTDinterwordspacing

\bibitem{jung:ieeecom:timevariant}
\BIBentryALTinterwordspacing
P.~Jung and G.~Wunder, ``{On Time-Variant Distortions in Multicarrier with
  Application to Frequency Offsets and Phase Noise},'' \emph{IEEE Trans. on
  Communications}, vol.~53, no.~9, pp. 1561--1570, Sep. 2005. [Online].
  Available:
  \url{http://ieeexplore.ieee.org/xpls/abs\_all.jsp?arnumber=1287432}
\BIBentrySTDinterwordspacing

\bibitem{ronshen:duality}
A.~Ron and Z.~Shen, ``{Weyl--Heisenberg frames and Riesz bases in
  L\_2(R\^{}d)},'' \emph{Duke Math. J.}, vol.~89, no.~2, pp. 237--282, 1997.

\bibitem{daubechies:tenlectures}
I.~Daubechies, ``{Ten Lectures on Wavelets},'' \emph{Philadelphia, PA: SIAM},
  1992.

\bibitem{Prete:1999}
V.~D. Prete, ``{Estimates, decay properties, and computation of the dual
  function for Gabor frames},'' \emph{Journal of Fourier Analysis and
  Applications}, 1999.

\bibitem{Laugesen2009}
R.~S. Laugesen, ``{Gabor dual spline windows},'' \emph{Applied and
  Computational Harmonic Analysis}, vol.~27, no.~2, pp. 180--194, Sep. 2009.

\bibitem{Christensen:2012}
O.~Christensen, H.~O. Kim, and R.~Y. Kim, ``{Gabor windows supported on [-1,1]
  and dual windows with small support},'' \emph{Advances in Computational
  Mathematics}, vol.~36, no.~4, pp. 525--545, 2012.

\bibitem{Sondergaard12}
P.~Sondergaard, ``{Efficient Algorithms for the Discrete Gabor Transform with a
  Long FIR Window},'' \emph{Journal of Fourier Analysis and Applications},
  vol.~18, no.~3, pp. 456--470, 2012.

\bibitem{vangelista2001efficient}
L.~Vangelista and N.~Laurenti, ``{Efficient implementations and alternative
  architectures for OFDM-OQAM systems},'' \emph{IEEE Transactions on
  Communications}, vol.~49, no.~4, pp. 664--675, 2001.

\bibitem{chu1972polyphase}
D.~Chu, ``{Polyphase codes with good periodic correlation properties
  (corresp.)},'' \emph{IEEE Transactions on Information Theory}, vol.~18,
  no.~4, pp. 531--532, 1972.

\bibitem{Bol01}
H.~B\"{o}lcskei, ``{Blind estimation of symbol timing and carrier frequency
  offset in wireless OFDM systems},'' \emph{IEEE Transactions on
  Communications}, vol.~49, no.~6, pp. 988--999, 2001.

\end{thebibliography}

\end{document}